\documentclass[pdftex,nofootinbib,superscriptaddress,aps,prfluids,notitlepage,longbibliography]{revtex4-2}
\usepackage{multirow,enumerate,epsfig,graphics,amssymb,amsmath,subeqnarray,mathrsfs,color,xcolor,epstopdf}
\usepackage{natbib}
\usepackage{color,epsfig,graphics,amssymb,mathcmd,amsmath,subeqnarray,graphicx,amsthm,mathrsfs} 
\usepackage[colorlinks=true, citecolor=red, linkcolor=blue ]{hyperref}
\usepackage{mathtools,bm} 
\usepackage{rotating,mathrsfs}
\usepackage{booktabs,subcaption,amsfonts,dcolumn}
\usepackage{multirow}
\usepackage[normalem]{ulem}
\usepackage{xcolor}

\makeatletter
\def\sgn{\mathop{\operator@font sgn}}
\def\threevdots{\vbox{\baselineskip1\p@ \lineskiplimit\z@
  \kern6\p@\hbox{.}\hbox{.}\hbox{.}}}
\makeatother

\begin{document} 
\title{Data-driven modelling for drop size distributions} 
\author{T. Traverso}
\affiliation{The Alan Turing Institute, British Library, 96 Euston Road, London NW1 2DB, UK}
\affiliation{Department of Aeronautics, Imperial College London, South Kensington Campus, London SW72AZ, UK}
\author{T. Abadie}
\affiliation{School of Chemical Engineering, University of Birmingham, Birmingham B15 2TT, UK}
\affiliation{Department of Chemical Engineering, Imperial College London, South Kensington Campus, London SW72AZ, UK}
\author{O. K. Matar}
\affiliation{Department of Chemical Engineering, Imperial College London, South Kensington Campus, London SW72AZ, UK}
\affiliation{The Alan Turing Institute, British Library, 96 Euston Road, London NW1 2DB, UK}
\author{L. Magri}
\email{l.magri@imperial.ac.uk}
\affiliation{Department of Aeronautics, Imperial College London, South Kensington Campus, London SW72AZ, UK}
\affiliation{The Alan Turing Institute, British Library, 96 Euston Road, London NW1 2DB, UK}
\begin{abstract}
 The prediction of the drop size distribution (DSD) resulting from liquid atomization is key to the optimization of multi-phase flows, from gas-turbine propulsion, through agriculture, to healthcare.  Obtaining high-fidelity data of liquid atomization, either experimentally or numerically, is expensive, which makes the exploration of the design space difficult. First, to tackle these challenges, we propose a framework to predict the DSD of a liquid spray based on data as a function of the spray angle, the Reynolds number, and the Weber number. 
 Second, to guide the design of liquid atomizers, the model  accurately predicts the volume of fluid contained in drops of specific sizes whilst providing uncertainty estimation. 
 To do so, we propose a Gaussian process regression (GPR) model, which infers the DSD and its uncertainty form the knowledge of its integrals, and of its first moment, i.e., the mean drop diameter. 
 Third, we deploy multiple GPR models to estimate these quantities at arbitrary points of the design space from data obtained from a large number of numerical simulations of a flat fan spray. The kernel used for reconstructing the DSD incorporates prior physical knowledge, which enables the prediction of sharply peaked and heavy-tailed distributions. 
 Fourth, we compare our method with a benchmark approach, which estimates the DSD by interpolating the frequency polygon of the binned drops with a GPR. We show that our integral approach is significantly more accurate, especially in the tail of the distribution (i.e., large, rare drops), and it reduces the bias of the density estimator by up to ten times. 
 Finally, we discuss physical aspects of the model's predictions and interpret them against experimental results from the literature.  
 This work opens opportunities for modelling drop size distribution in multiphase flows from data.

\end{abstract}
\maketitle
\section{Introduction}  \label{sec1}
Liquid atomization is a phenomenon that appears in a large number of applications, from agriculture \cite{li_2021}, through drug delivery, to fuel injection and cosmetics \cite{Lefebvre_1995_ASME, lefebvre2017atomization}.
In most applications, it is desirable to have an {\it a priori} knowledge of the liquid dispersion structure, in particular, the distribution of droplet sizes as a function of the main control parameters and working conditions of the atomizer. In agriculture, the droplet size and velocity distributions are important to ensure accurate delivery and retention of pesticide on the target \cite{ellis1997} as the distribution of smaller drops determines the sensitivity of the coverage to environmental effects, such as wind \cite{miller1988}. A similar problem occurs in automotive paint sprays \cite{Babinsky2002}. In respirable sprays for medicinal applications, the width of the drop size distribution (DSD) is of vital importance because drops smaller than a critical size are ejected from the body during exhalation, whereas droplets that are too large remain trapped in the respiratory system far upstream of their intended target \cite{Babinsky2002}. In chemically reacting flows, before ignition, the DSD of atomized liquid fuels plays a central role in determining the thermal efficiency and the emissions of pollutants because it influences, e.g., the local evaporation rate of the fuel, the tendency of drops to be transported by the flow, and the spray penetration  \cite{Mohan_2013_combustion, schmehl2000cfd}.

The dispersion structure of the atomized fluid is determined by a large number of factors, such as the presence of boundaries \cite{Andreassi_2007,wang1995experimental}, the pressure drop across the nozzle or its shape \cite{wang1995experimental}, as well as the physical properties of the fluid itself,  to name a few. For this reason, although the mathematical analysis of the governing equations describing sprays and jets can help understand the fundamental mechanisms leading to drop formation, it can provide only qualitative prediction regarding the DSD \cite{villermaux2002life, Villermaux2007}. Thus, experimental investigation is often necessary to obtain quantitatively accurate information. Experiments provide high-fidelity data about the detailed shape of the DSD resulting from atomization; however, this is a low-throughput and expensive approach \cite{khopkar2012industrial_book}. An  alternative approach is to perform high-fidelity numerical simulations, which are a reliable tool to observe the details of the physical mechanisms at play in multiphase systems, e.g.,  \cite{constante2021direct,ling2017spray,shin2018hybrid}. However, high-fidelity simulations  also have limitations because they are typically computationally expensive simulations, which make the systematic exploration of large design spaces difficult. 

To mitigate the aforementioned challenges, Sacks et al. \cite{sacks1989design} modelled the deterministic output of numerical simulations as a stochastic process, providing the statistical basis for the optimal design of experiments (DOE), or active learning problem. In this framework, the goal is to choose the input sample carefully so that  the information for the output quantity of interest (QoI) is maximized \cite{Chaloner_1995}. Recently, a similar framework was developed to design experiments using an output-weighted metric, with the objective of achieving the fastest possible convergence for the output statistics rather than  reducing the accuracy in regions of the design space of little interest, with applications to rare events (e.g., waves) \cite{sapsis2020output}. 


By modelling the output QoI of numerical simulations as a stochastic process it is possible to build a surrogate model (SM) of such QoI. In essence, a data-driven SM is a supervised learning algorithm that uses data to learn the mapping between points in the design (or input) space and the QoI. These SMs are sometimes referred to as \emph{statistical emulators} as they \emph{emulate} the solution of a physics-based model. The latter, which is numerically solved to obtain data, is thus called the \emph{simulator} \cite{Conti2010}. 

In the simplest case, the QoI is a scalar function of the input, e.g., the power output of a wind farm as a function of its layout and wind conditions \cite{bempedelis2023bayesian}. However, if the QoI is the probability $p(d)$ of finding a drop with diameter $d$, then it becomes a function of the internal coordinate $d$. This intricate QoI cannot be easily modelled as a scalar stochastic process. To address this, a possible solution is to use multiple SMs, with each SM learning the function at a specific point of the internal coordinates. An illustrative example is provided by \cite{Conti2010}, where the QoI is represented by the time series of a scalar quantity of interest, specifically the carbon concentration within forests. Their approach involves training a vector-valued Gaussian process regressions (GPR) model, in which each element of the output vector represents the carbon concentration at a given time. Alternatively, the output vector could represent the coefficients of a set of basis functions, e.g., spectral decomposition \cite{campbell2006sensitivity, Bayarri_2007_MS}, instead of the point-evaluation of the QoI at different locations. Nonetheless, in all these cases the QoI is reconstructed from the knowledge of some \emph{features}, such as its value at a set of points in time, or its spectral representation.

In this work we propose a SM to learn the mapping from the working conditions of a flat-fan liquid spray to the resulting DSD $p(d)$, which is the QoI. The working conditions are parametrized by (i) the spray angle, (ii) the Reynolds and (iii) the Weber numbers of the jet. 
To do so, we exploit the ability of GPR models to reconstruct the unknown function by using observations of any linear functional of it, and not only of the point-evaluation functional. Other works exploited this idea in different contexts. For example, Longi et al.~\cite{Longi2020} used integral observations to optimize sensor placement, and Law at al. \cite{law2018variational} to model the spatial distribution of malaria incidence from aggregate data, while Solak et al. \cite{solak2002derivative} used derivative observations to model dynamical systems. In this work, we propose an algorithm that reconstructs the DSD by combining observations of integrals as well as of its first moment. 
These types of observations are relevant to multi-phase systems, and especially to atomization problems, because the integrals of the DSD between two points $d_1$ and $d_2$ is directly related to the volume of atomized fluid contained in droplets with size $d_1<d<d_2$. At the same time, by imposing the first moment of the DSD, we significantly reduce the bias of the estimator. The performance of such density estimators are compared to the case where point-wise evaluation of the DSD are used for its reconstruction, showing significant improvement. 
Finally, we remark how the proposed approach does not rely on \emph{a priori} parametrization of the DSD in terms of known distributions (e.g., Gamma or log-normal), as done in the literature when reconstructing DSD from data, e.g., \cite{Villermaux2007, cao2010polarimetric, tengeleng2014performance, Kooij2018}. The prior knowledge on the typical shape of DSD is  embedded in the choice of the kernel, which is a key aspect of the model selection.

The paper is structured as follows. In section \ref{sec2}, we introduce the formalism of GPR adapted to include any observation that can be expressed as a linear functional of $p(d)$. Specifically, the functionals will be integrals of $p$, their first moments. For the purpose of training and testing the SM, a data set of numerical simulations of the spray is produced at different working conditions. In section \ref{sec:GPR_pdf}, we introduce the computational fluid dynamics framework used to perform these simulations and how the continuous DSD and its uncertainty are reconstructed from the populations of drops. In section \ref{sec4}, the SM mapping the design parameters to the associated DSD is described, the performance is  tested, and the outputs are interpreted from a physical perspective. Finally, concluding remarks are provided in section \ref{sec:concl}.


%

\section{Gaussian process regression with functional observations} \label{sec2}
The goal of this section is to introduce the formalism of GPR by considering the general case where inference is based on observations of linear functionals $L[f]$ of the unknown function estimated by the Gaussian process, $f$. This includes the case where $L$ is the point-evaluation functional $L_x[f]=f(x)$, that is when information about $f$ is provided by some samples of its value at discrete points.
\subsection{Functions as Gaussian processes} \label{sec:GPR1.1}

   A Gaussian process (GP) is a collection of random
variables that follow a joint Gaussian distribution. In the context of GPR, the random value represents the value of a function $f(\textbf{x}):\mathcal{X}\in\mathbb{R}^{n} \mapsto \mathbb{R}$ at point $\textbf{x}$ \citep{Rasmussen2006}. 
If we let $\textbf{X}=\{\textbf{x}_i\}_{i=1}^N$ be any finite set of points in $\mathcal{X}$, the vector $f(\textbf{X})=[f(\textbf{x}_1),\dots,f(\textbf{x}_N)]^{\textrm{T}}$ is modelled as a GP and it is denoted by
\begin{equation}
    f(\textbf{X}) \sim  \mathcal{GP}(m(\textbf{X}),  k(\textbf{X}, \textbf{X})  ) , \label{prior}
\end{equation}
where $m(\textbf{X})$ is the $N$-dimensional vector with elements $m(\textbf{x}_i)= \mathbb{E}\left[ f(\textbf{x}_i) \right]$, i.e., the mean of $f$ at $\textbf{x}_i$, and $k(\textbf{X}, \textbf{X})$ is the $N$-by-$N$ covariance matrix defined as
\begin{align}
      k_{ij} = k(\textbf{x}_i,\textbf{x}_j) = \mathbb{E}\left[ \left( f(\textbf{x}_i) - m(\textbf{x}_i) \right) \left( f(\textbf{x}_j) - m(\textbf{x}_j) \right) \right]. \label{kij}
\end{align}
The GP is defined by the knowledge of its mean and covariance functions.

Before the actual regression is performed (i.e., before data are considered), we chose the functions $m$ and $k$ based on prior knowledge about the unknown function $f$. In this work, such prior knowledge is embedded in the form of the kernel, $k(\textbf{x},\textbf{x}'|\boldsymbol{\theta})$ (reported in equations \eqref{kernel} and \eqref{ARD_SE} below), which is chosen {\it a priori} and tuned by adjusting 
the hyperparameters $\boldsymbol{\theta}$. As more rigorously discussed in sections \ref{sec:features} and \ref{sec:surr_mod_str}, these hyperparameters quantify how \emph{fast} the function $f$ changes as we move in the input space. This training is known as \textit{type II maximum likelihood approximation} or ML-II \cite{Rasmussen2006}. 
The prior mean function is assumed to be zero, that is $m(\textbf{x})=0$ for all $\textbf{x}$. This assumption preserves the generality of the method as it can always be enforced by shifting the data set by its mean.

Let us now consider a family of linear functionals of $f$, each denoted by $L_i[f]:\mathcal{F} \mapsto \mathbb{R}$ and such that $L_i[af+g] = aL_i[f] + L_i[g]$ for $f,g \in \mathcal{F}$, where $\mathcal{F}$ is a reproducing kernel Hilbert space \citep{Alvarez2012}.
We consider that we have a noisy measurement of each $L_i$, denoted by $y_i=L_i[f] + \epsilon_i$, where $\epsilon_i$ is the zero-mean, additive, independent, and identically distributed Gaussian noise corrupting the observed value of $L_i$, and whose variance is denoted by $\mathbb{V}[\epsilon_i]$.
 As discussed in section \ref{sec:features}, these functional observations of $f$ are definite integrals and the first moment of $f$. Here, the unknown function is the DSD of a liquid spray, $f=p(d)$, and its observed values, $y_i$, will thus represent, e.g., the probability to find a droplet whose size is within a certain range (i.e., the extremes of integration defining $L_i$). 
%

Considering the assumptions on $\epsilon_i$, and that linear transformations of GPs are also GPs \cite{OHagan1991BayesHermiteQ}, the prior distribution for the $i^{\rm th}$ observation $y_i$ is also Gaussian:
\begin{equation}
    y_i = \mathcal{N} \left( L_i[m], L_i^2[k] + \mathbb{V}[\epsilon_i] \right) , \label{yi_Gauss}
\end{equation}
where $L_{i}^2[k]$ is expressed by
\begin{equation}
    L_{i}^2[k] = L_i\left[ L_i\left[ k(\cdot,\textbf{x}') \right] \right] = L_i\left[ L_i\left[ k(\textbf{x}, \cdot) \right] \right].  \label{Lii}
\end{equation}
If the commonly-used {point-evaluation} functional $L_i=L_{\textbf{x}_i}[f]=f(\textbf{x}_i)$ is employed, then equation \eqref{yi_Gauss} becomes  $y_i = \mathcal{N} \left(m(\textbf{x}_i), k(\textbf{x}_i,\textbf{x}_i) + \mathbb{V}[\epsilon_i] \right)$ \cite{Rasmussen2006}. 
As equation \eqref{yi_Gauss} shows, a GP distribution on $f$ gives a Gaussian distribution on any linear functional of $f$. This fact, together with the framework introduced in sections \ref{sec:GPR1.2} and \ref{sec:GPR1.3}, can be used to infer the integral of $f$ from the knowledge of its value at a limited number of locations. This procedure, known as \emph{Bayesian quadrature}, treats the problem of numerical integration as one of statistical inference \cite{OHagan1991BayesHermiteQ} where the main source of uncertainty arises because it is not affordable to compute the function $f$ at every location \cite{rasmussen2003bayesian, gunter2014sampling}, e.g., when integrating functions over high-dimensional domain. In this work, we exploit this idea to perform the inverse task, that is, to infer $f$ from the partial knowledge of some linear functionals of it (e.g., integrals). 

\subsection{Selecting a Gaussian process model: training the hyperparameters} \label{sec:GPR1.2} 
 Let us consider the regression problem of estimating the value of $\textbf{f}_* = f(\textbf{X}_*)$ at some test locations $\textbf{X}_* = \{\textbf{x}_{i}\}_{i=1}^{N_*}$ from the knowledge of $N_f$ features (i.e., linear functionals) of $f$. 
The definition of these functionals along with their observed values forms the data set $\mathcal{D}=\{ (L_i,y_i)\}_{i=1}^{N_f}$. For brevity, we define the collection of all the functionals by $\textbf{L} = \{L_i \}_{i=1}^{N_f}$. Each $L_i$ is associated with its observed value, collected in the vector $\textbf{y}=[y_1,\dots,y_{N_f}]^{\textrm{T}}$. This vector of observations is, in turn, associated with the noise vector, $\boldsymbol{\epsilon} = [\epsilon_1,\dots, \epsilon_{N_f}]^{\textrm{T}}$, and with the $N_f$-by-$N_f$ diagonal matrix containing the variance of such noise along the main diagonal (i.e., uncorrelated noise), $\mathbb{V}[\boldsymbol{\epsilon}]$. 

The GPR comprises two main steps in which the data set $\mathcal{D}$ is used to (i) select a model, and (ii) condition the model on the observations, respectively. The model selection consists of optimizing the hyperparameters $\boldsymbol{\theta}$ defining the covariance function $k=k(\textbf{x},\textbf{x}'|\boldsymbol{\theta})$ based on the observed data, which is the \emph{training}.  This is achieved by maximizing the marginal likelyhood 
$\mathcal{L}(\boldsymbol{\theta}) = p(\textbf{y}|\textbf{L},\boldsymbol{\theta})$ given by \cite{Rasmussen2006} 
\begin{equation}
    \log p(\textbf{y}|\textbf{L},\boldsymbol{\theta}) = -\frac{1}{2} \textbf{y}^{\textrm{T}} k_y^{-1} \textbf{y}  - \frac{1}{2}\log |k_y| - \frac{N_f}{2} \log 2\pi, \label{LL}
\end{equation}
where $k_y = k_L(\textbf{L}, \textbf{L}) + \mathbb{V}[\boldsymbol{\epsilon}]$ is the covariance between the noisy observations of $\textbf{L}$. The $N_f$-by-$N_f$ covariance matrix of the noise-free observations, $k_L(\textbf{L}, 
\textbf{L})= k_{L,ij} $, is obtained by exploiting the bilinearity of covariance \cite{rice2006mathematical}, yielding
\begin{align}
      k_{L,ij} = L_i\left[ L_j \left[ k(\cdot ,\cdot) \right] \right] = L_j\left[ L_i \left[ k(\cdot ,\cdot) \right] \right]. \label{kL}
\end{align}
The definition of the functionals used in this work and the associated covariance terms in equations \eqref{kL} are explicitly reported in section \ref{sec:features}. 
Once the model is trained, we denote the optimized covariance function by $\hat{k}=k(\textbf{x}, \textbf{x}' | \boldsymbol{\hat{\theta}})$, where $\hat{\theta}$ corresponds to 
\begin{equation}
  \hat{\boldsymbol{\theta}} \equiv \arg \max_{\boldsymbol{\theta}} p(\textbf{y}|\textbf{L},\boldsymbol{\theta}),  \label{training} 
\end{equation}
and similarly for the covariance matrix between observations of different functionals, $\hat{k}_{L,ij} = L_i\left[ L_j \left[ \hat{k}(\cdot ,\cdot) \right] \right]$.

\subsection{Posterior distribution} \label{sec:GPR1.3}
Once the model is trained, the next step is to condition the prior, see Eq.~\eqref{prior}, on the data, $\mathcal{D}$. In practice,  that is to find the posterior distribution for (i.e., prediction of)  the value of $f$ at the test locations, $\textbf{f}_*$. These predictive equations for $\textbf{f}_*$ \citep{Rasmussen2006} are here generalized to observations of functionals other than the point-evaluation functional. They read
\begin{align}
    \textbf{f}_*|\mathcal{D},\textbf{X}_* &\sim \mathcal{N}( \mathbb{E}[\textbf{f}_*] , \textrm{cov}(\textbf{f}_*)  ), \ \textrm{where}  \label{eq:pred1} \\
    \mathbb{E}[\textbf{f}_*]   &= \hat{k}_L(\textbf{L}_*,\textbf{L}) \hat{k}_y^{-1} \textbf{y}, \label{eq:pred2} \\
    \textrm{cov}(\textbf{f}_*) &=  \hat{k}_L(\textbf{L}_*,\textbf{L}_*) - \hat{k}_L(\textbf{L}_*,\textbf{L}) \hat{k}_y^{-1}  \hat{k}_L(\textbf{L},\textbf{L}_*) , \label{eq:pred3}
\end{align} 
where $\hat{k}_y = \hat{k}_L(\textbf{L}, \textbf{L}) + \mathbb{V}[\boldsymbol{\epsilon}]$, and where $\textbf{L}_* = \{ L_{\textbf{x}}[f]  | \textbf{x} \in \textbf{X}_*  \}$ denotes the set of point-evaluation functionals at the test locations $\textbf{X}_*$. In other words, in equation~\eqref{eq:pred2} we now use the optimized covariance ($\hat{k}_y$) and the data ($\textbf{y}$) to predict the most likely value of the unknown function $f$ at a set of test locations ($\textbf{X}_*$) in which we have no data, i.e., we compute $\mathbb{E}[\textbf{f}_*]=\mathbb{E}[f(\textbf{X}_*)]$. Equation~\eqref{eq:pred3} is employed to compute the covariance between these predictions, $\textrm{cov}(\textbf{f}_*)$. Its diagonal represents the variance of the prediction at each test point, $\textbf{X}_*$, thus quantifying its accuracy, and it is independent of the value of the observations, $\textbf{y}$. As discussed in section \ref{sec4}, this fact is exploited to optimally design part of the numerical experiments because it is possible to know how a new experiment affects the uncertainty of the prediction before it is run.

In sections \ref{sec:features} and \ref{sec:results} we use the predictive equations \eqref{eq:pred1}-\eqref{eq:pred3} to predict the DSD of a liquid spray, $f=p(d)$, that is, the probability of finding a spherical drop with diameter $d$. We do so by using 
a specific form of the covariance function that well encodes our prior knowledge about the general characteristics of DSD (e.g., heavy tail distributions). Finally, in section \ref{sec4}, we model each of these features of $p$ using as many GPR models, effectively treating them as unknown functions of the  working conditions of the spray, which is the input space of the SM. By doing so, we obtain a SM that maps points in the space of the working conditions of the sprays to the associated DSD and its uncertainty.

\section{GPR for probability density estimation for sprays} 
\label{sec:GPR_pdf}
 
In this section, we propose a novel application of the framework explained in Sec.~\ref{sec2} in which observations include (i) definite integrals and (ii) the first moments of the unknown function. The unknown function represents the unobservable underlying probability density (PDF) that generates a finite data sample, that is the data. 
Each element of the sample, or population, corresponds to the diameter of a drop within a liquid spray. To obtain the population of drops, a flat-fan liquid spray is modelled by means of numerical simulations, as described in section \ref{sec:CFD}. In section \ref{sec:features}, we then discuss how the GPR framework is used to estimate the underlying PDF generating the DSD. Finally, the results are discussed in section \ref{sec:results}. 

\subsection{Numerical simulations of a liquid jet} \label{sec:CFD} 
%

\subsubsection{Governing equations and dimensionless parameters} \label{sec:nondim} 

Atomization is the physical process by which an initial liquid, here a liquid sheet, is fragmented into smaller droplets due to a cascade of events. It is characterized by an initial phase, when linear instabilities of the sheet's free surface saturate to form ligaments. In a second phase, ligaments breakup into a number of smaller drops, which are responsible for the shape of the DSD of the spray  \citep{Villermaux2007}. These phases are referred to as {\textit primary} and \textit{secondary} atomization, respectively. 
The dynamics of an incompressible, Newtonian liquid jet are described by the one-fluid formulation of the incompressible Navier-Stokes equations, which under the assumptions of isothermal flow, and in the absence of interfacial mass transfer, are expressed by
    \begin{equation}
    \boldsymbol{\nabla \cdot {u}} = 0, 
    \label{eq:continuity}
    \end{equation}
    \begin{equation}
    \rho\left(\frac{\partial \boldsymbol{u}}{\partial t}+\boldsymbol{u} \cdot \boldsymbol{\nabla u}\right)=-\boldsymbol{\nabla}  p+\boldsymbol{\nabla} \cdot \left( \eta \left[ \boldsymbol{\nabla u} + \boldsymbol{\nabla u^T }\right] \right)+\boldsymbol{F_\sigma} + \boldsymbol{g}, 
    \label{eq:NS}
    \end{equation}
where $\rho$, $\eta$ and $\boldsymbol{u}$ are the local density, dynamic viscosity, and velocity, respectively; $\boldsymbol{g}$ is the acceleration due to gravity, and the $\boldsymbol{F_\sigma}$ is the capillary force, which can be written as $\boldsymbol{F_\sigma} = \sigma \kappa \mathbf{n} \delta_{s}$, with $\sigma$, $\boldsymbol{n}$, $\kappa$, and $\delta_s$ denoting the surface tension (considered constant), the unit normal to the interface, the local curvature of the interface, and the Dirac delta distribution associated with the interface, respectively. 
We consider an ellipsoidal nozzle with minor and major semi-axis $r_a$ and $r_b$, respectively. Using twice the nozzle minor semi-axes $R_0$, twice the jet maximum velocity $U_0$, $t_c = R_0/U_0$, and $\rho_l U_0^2$ as the characteristic length, time, velocity, and pressure, respectively, and the liquid density $\rho_l$ and viscosity $\eta_l$ to scale the density and viscosity, respectively, the equations above can be written as \cite{abadie2015,popinet2018_review}
    \begin{equation}
    \boldsymbol{\Tilde{\nabla}} \cdot \boldsymbol{\Tilde{u}} = 0, 
    \label{eq:continuity}
    \end{equation}
\begin{equation}
     \Tilde{\rho} \left(\dfrac{\partial \boldsymbol{\Tilde{u}}}{\partial \Tilde{t}} +  \boldsymbol{\Tilde{u}} \cdot \Tilde{\boldsymbol{\nabla}}\boldsymbol{\Tilde{u}} \right)
  = - \Tilde{\boldsymbol{\nabla}} \Tilde{p} + \dfrac{1}{Re} \Tilde{\boldsymbol{\nabla}}\cdot \left(\Tilde{\eta} \left[ \boldsymbol{\Tilde{\nabla} \Tilde{u}} + \boldsymbol{\Tilde{\nabla} \Tilde{u}^T}\right] \right)
  + \dfrac{1}{We} \Tilde{\kappa} \Tilde{\delta_s} \boldsymbol{n} + \dfrac{1}{Bo} \boldsymbol{\Tilde{g}},
    \label{eq:scaled_NS}
\end{equation}
where the tilde denotes dimensionless quantities. Considering flows where the capillary forces dominate the acceleration due to gravity ($Bo=\rho_l g R^2_0/\sigma \ll 1$), the two dimensionless parameters that influence the spray dynamics, thus the resulting droplet size distributions, are the Reynolds number $Re = \rho_l U_0 R_0 / \eta_l$ and the Weber number $We = \rho_l U_0^2 R_0 / \sigma$.

\subsubsection{Numerical methods} \label{sec:NumMethFl} 
\begin{figure}[!htb] 
\centering
  \subcaptionbox{}{\includegraphics[width=0.48\textwidth]{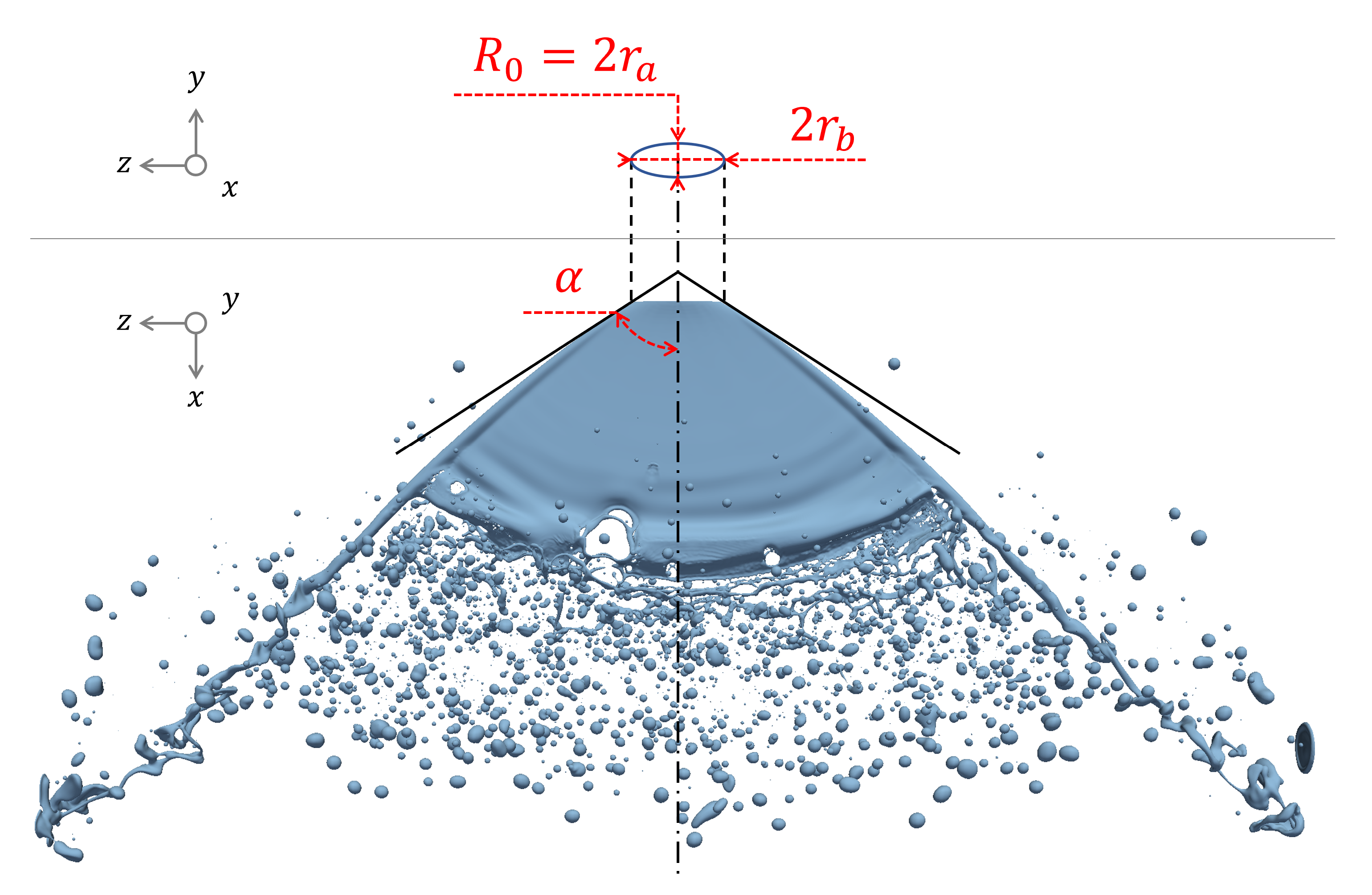}} 
  \subcaptionbox{}{\includegraphics[width=0.48\textwidth]{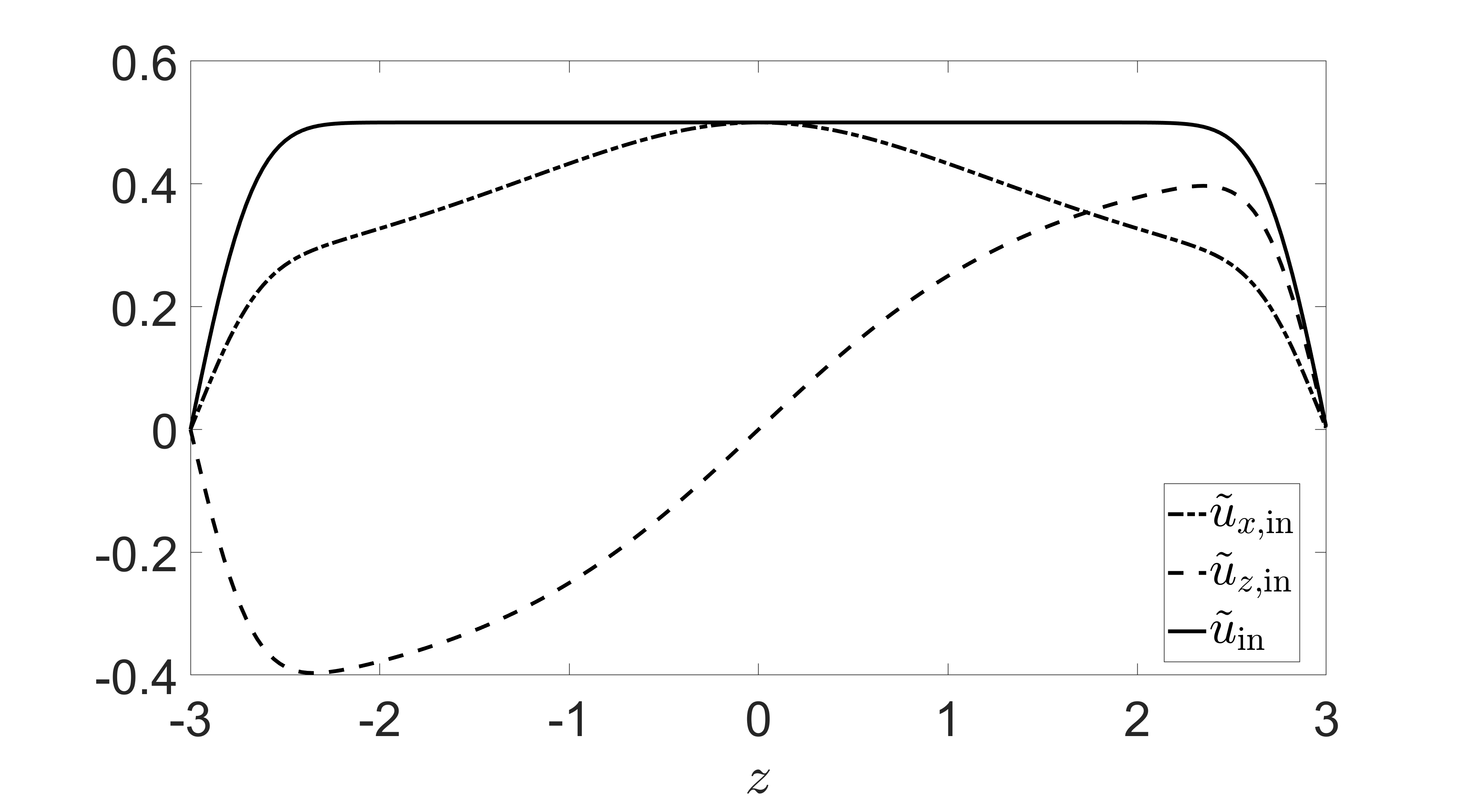}}  
  \subcaptionbox{}{\includegraphics[width=0.48\textwidth]{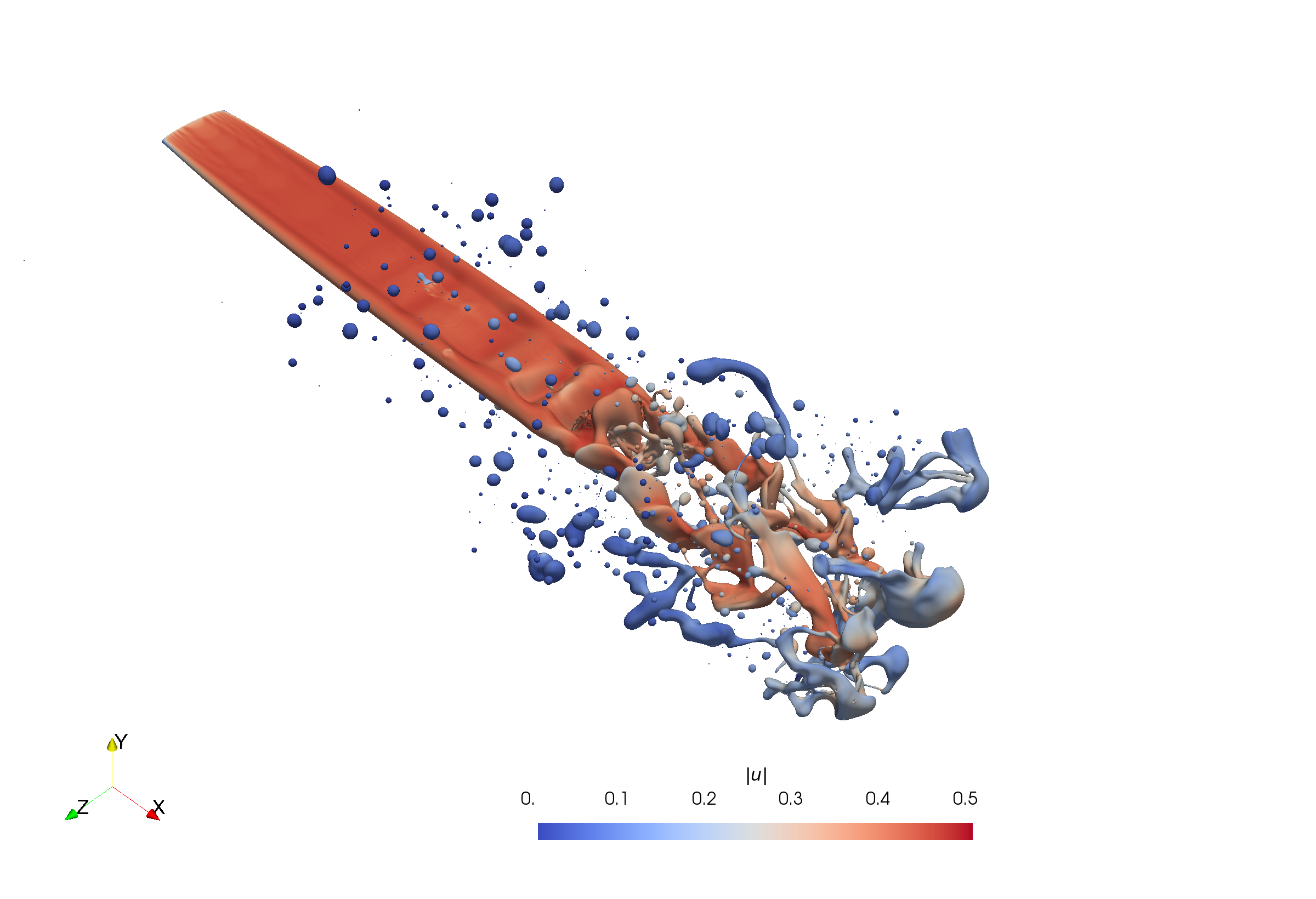}}
  \subcaptionbox{}{\includegraphics[width=0.48\textwidth]{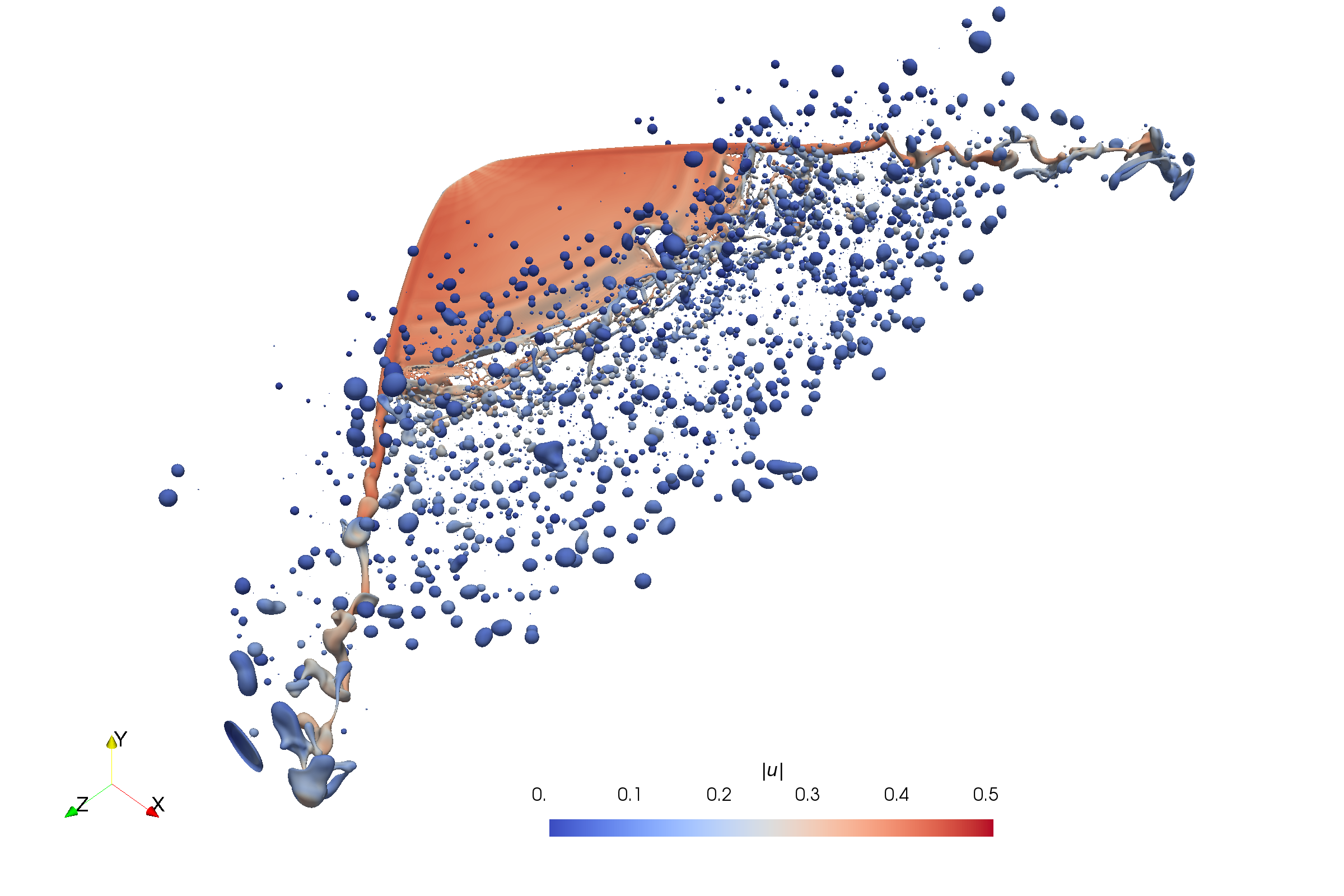}} 
  \subcaptionbox{}{\includegraphics[width=0.4\textwidth]{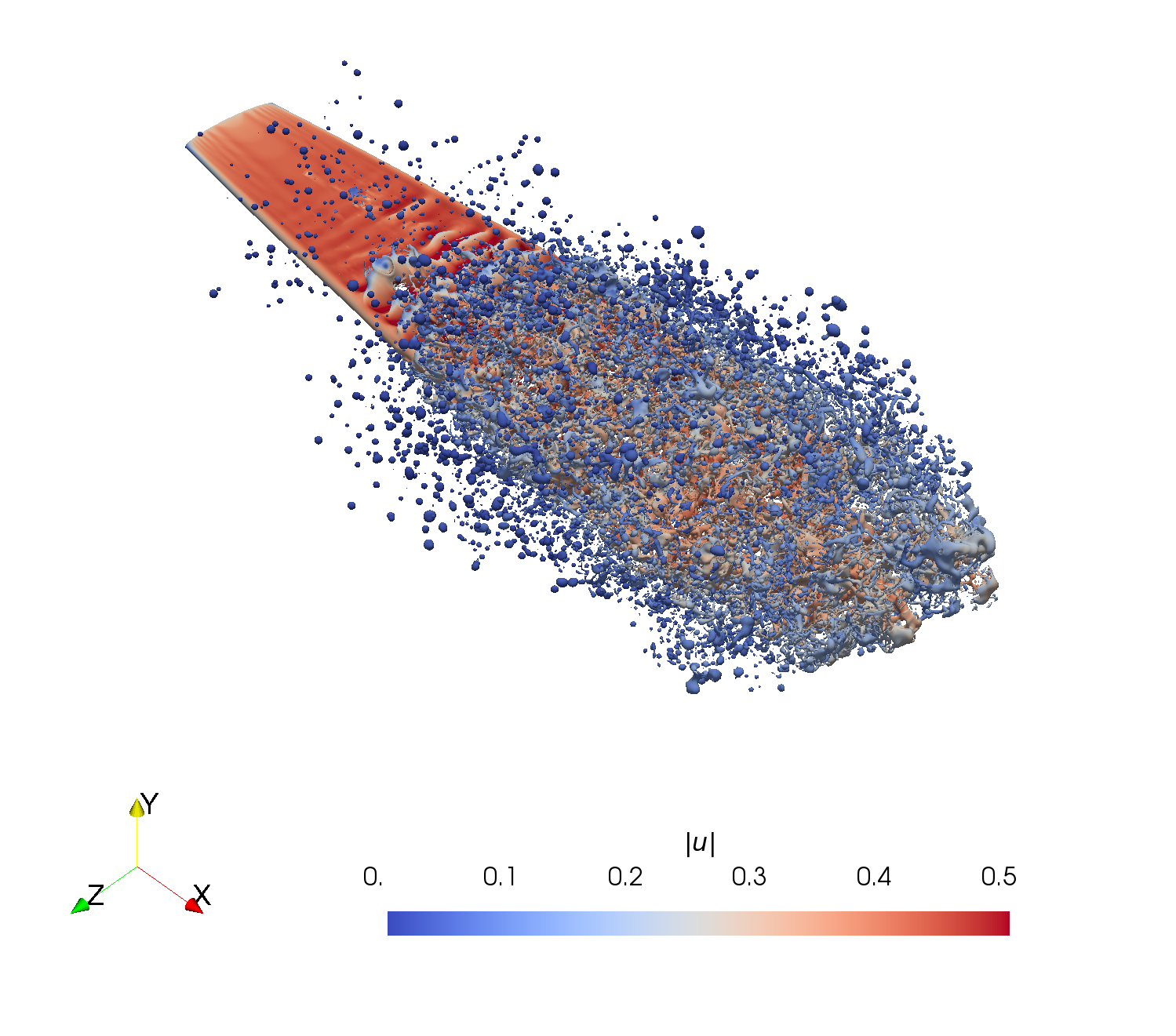}}
  \subcaptionbox{}{\includegraphics[width=0.4\textwidth]{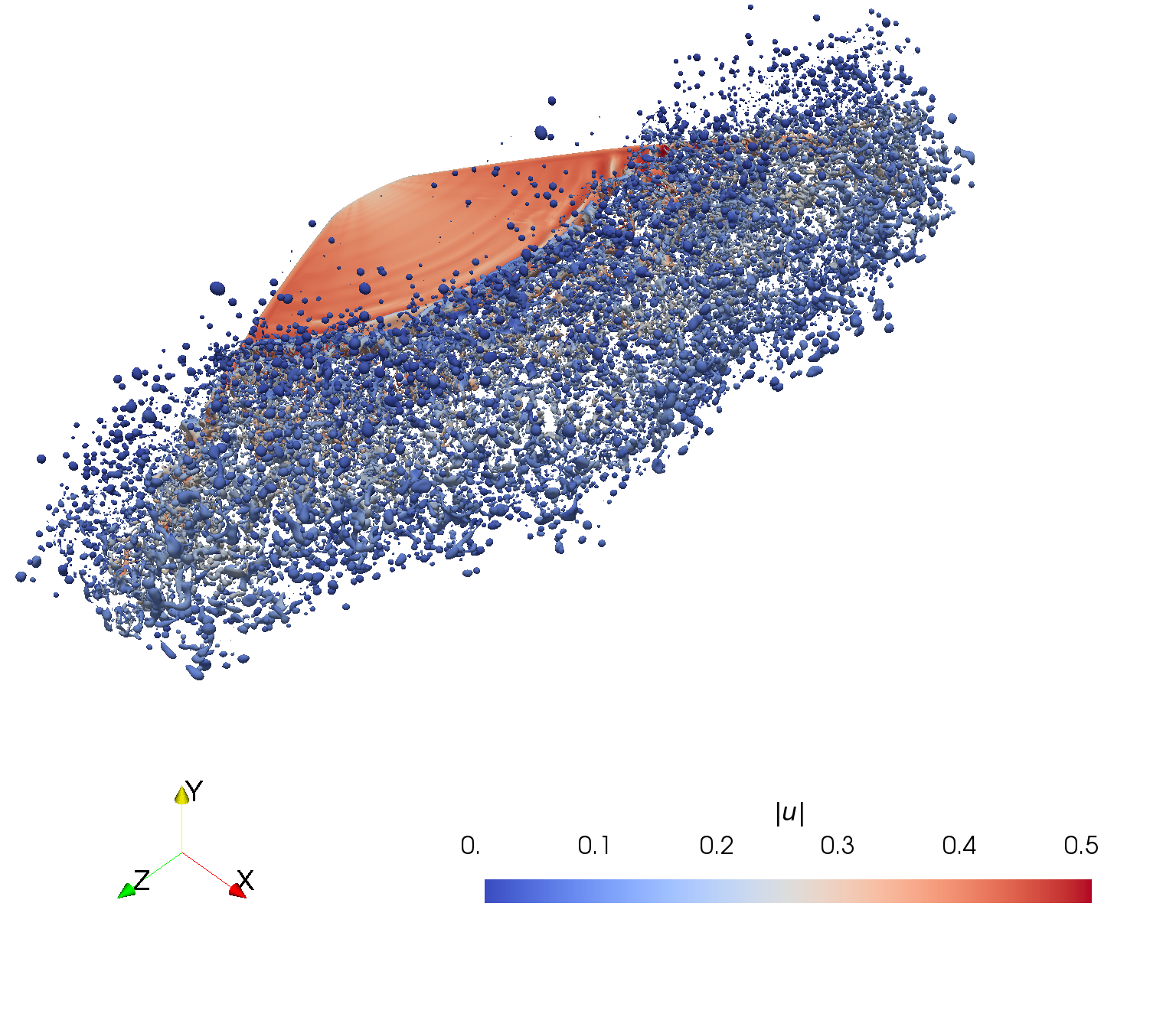}}
  \caption{(a) Qualitative representation of the spray angle and the ellipsoidal region (through orthographic projection) letting the inflow into the domain. (b) Axial ($\Tilde{u}_{x, {\rm in}}$), lateral ($\Tilde{u}_{z, {\rm in}}$), and total inflow velocity, $\boldsymbol{\Tilde{u}}_{{\rm in}}=(\Tilde{u}_{x, {\rm in}}(z), 0, \Tilde{u}_{z, {\rm in}}(z))$, evaluated along the main diagonal of the ellipse, i.e., $x=0$, $y=0$, and $z\in [-r_b/2 , r_b/2]$ with $r_b=6$. 
  Values relative to a spray angle of $60^{\circ}$, which corresponds to $\tan^{-1}(\Tilde{u}_{z, {\rm in}}/\Tilde{u}_{x, {\rm in}})$ as $z \rightarrow 0$. Fluid interface and velocity magnitude (color) of four sprays with (c) $Re=20$, $We=9$, $\alpha=10^{\circ}$, (d) $Re=20$, $We=9$, $\alpha=65^{\circ}$, (e) $Re=59$, $We=45$, $\alpha=10^{\circ}$, and (f) $Re=59$, $We=45$, $\alpha=65^{\circ}$. }  \label{fig:AMR}
\end{figure}

We perform numerical simulations of a liquid jet using the \emph{Basilisk} GPL-licensed computational code \cite{popinet2015_basilisk}. The Navier-Stokes equations are solved using the projection method \cite{popinet2009}, while the volume-of-fluid (VOF) method is used to capture and reconstruct the interface \cite{tryggvason2011, aniszewski2014_VOF}. Within this framework, the volume fraction $C$, which equals unity in the liquid phase 
and zero in the ambient phase, is passively advected with the flow:
\begin{equation}
    \frac{\partial {\tilde{\mathbf{u}}}}{\partial \tilde{t}}+\tilde{\mathbf{u}}\cdot \boldsymbol{\nabla} C=0.
\end{equation}
The local fluid viscosity $\eta$ and density $\rho$ are obtained by linearly averaging the liquid and gas properties with $C$
\begin{equation}
\tilde{\eta}=\frac{\eta_l C + (1-C)\eta_g}{\eta_l}, ~~~~~
\tilde{\rho}=\frac{\rho_l C + (1-C)\rho_g}{\rho_l}
\end{equation}
where $\rho_g$ and $\eta_g$ represent the gas density and viscosity, respectively, and $\rho_l/\rho_g = \eta_l/\eta_g = 10$ throughout. 
Surface tension forces are computed from the local normal and curvature of the interface, which are estimated from $C$ using the Height-Functions method \cite{afkhami2008_Curvature,popinet2009}. 

The simulations are performed on a cube with nondimensional linear size $\Tilde{L}=88$. On each side, we impose reflective boundary conditions (i.e., zero normal derivative) for all the state variables exceptions made for the pressure on the opposite side of the nozzle, which is set to zero. The inflow velocity, $\boldsymbol{\Tilde{u}}_{{\rm in}}=(\Tilde{u}_{x, {\rm in}}(z), 0, \Tilde{u}_{z, {\rm in}}(z))$, is imposed on an ellipsoidal region on the plane $x=0$ with major axes $r_b$ along direction $z$, and minor axes $r_a$ along direction $y$ (aspect ratio $r_b/r_a=6$), see figure \ref{fig:AMR} (a). The components of $\boldsymbol{\Tilde{u}}_{{\rm in}}$ are plotted in figure \ref{fig:AMR} (b); they are defined such that at each point inside the ellipse  
\begin{equation}
    \begin{cases}
     2\sqrt{ \Tilde{u}^2_{x, {\rm in}} + \Tilde{u}^2_{z, {\rm in}} }  = U_0 {\rm erf} \left[ \frac{2(r_b-|z|)}{\delta} \right]  \\
  \Tilde{u}_{z, {\rm in}} /  \Tilde{u}_{x, {\rm in}}  = z \tan{\alpha} / r_b
    \end{cases}\ ,
\end{equation}
which ensures that the desired spray angle $\alpha$ is obtained, and $\delta = r_b/4$. 

The solver employs local adaptive mesh refinement (AMR) \cite{puckett1992_AGR} using the octree mesh type. 
Basilisk’s procedure for local mesh adaptation employs the wavelet transform to assess the discretization error of a given scalar field. 
In our simulations, we adapt the mesh according to the error on the local non-dimensional fluid velocity $\boldsymbol{u}$, and the gradient of the volume fraction of the fluid in the jet $\boldsymbol{\nabla} C$, which indicates the vicinity of an interface. Throughout this study, the threshold error value is set to $0.1$, and the maximum level of refinement is set equal to $11$. Figures \ref{fig:AMR} (c) to (f) show four jets discretised by $O(10^7)$ grid points corresponding with different working conditions of the jet, i.e., different values of $\alpha$, $Re$, and $We$ (see caption). The simulations are concluded when a sufficient number of droplets has formed, typically $O(10^3)$, and can take up to $72$ hours on $24$ cores Intel i7-12700.

When the simulation is concluded only droplets with an equivalent diameter $d_i$ such that $0.08<d_i<0.8$ are considered (minimal cell size is $l_{\rm min}\approx0.04$). Then, non-spherical drops are excluded from the sample by applying a high-pass filter based on their sphericity $\phi_i = \pi^{1/3} (6V_i)^{2/3} / A_i$. To estimate the surface area of the $i^{\rm th}$ drop, $A_i$,  we first note that the vector field $\boldsymbol{\nabla} C$ approximates a Dirac delta function on the interface between the drop and the ambient fluid. Within a computational cell on such an interface, we define the vector $d\textbf{A}$ whose modulus  is equal to the surface area and its orientation is perpendicular to it. The area of such an interface is then given by $|d\textbf{A}| = ( dA_1^2 + dA_2^2 + dA_3^2)^{1/2}$, where $d A_j~(j=1,2,3)$ are expressed by
\begin{equation}
    dA_j = \int_{V_{cell}} \boldsymbol{\nabla} C_j {\rm d}V.
\end{equation}
Finally, $A_i$ is computed by summing all the infinitesimal contributions $dA_j$ relative to the $i^{\rm th}$ drop. 
Drops with $\phi_i<0.9$ are thus excluded as they are likely to be ligaments that would eventually break up into smaller drops, thus corrupting the statistics of the DSD. We observe that, on average, around $3\%$ of the total number of drops are filtered out based on sphericity, corresponding to $11\%$ of the total atomized volume. If the filter based on the size is also considered, the drops included in the final count are on average $74\%$ of the total, making up for $46\%$ of the total volume.  Analogous size-based band-pass and sphericity-based filters are applied also when using optical techniques to study atomization due to the intrinsic limits of the image acquisition system \citep{Nowak_2021, OxfordLasersSoftware, jin2018experimental}. 

\subsection{Features of the drop size distribution and their covariance} \label{sec:features} 

\subsubsection{Observations of the mean and cumulative probabilities} \label{sec:kLij}
The drop population obtained from the numerical simulation of the spray is treated as an independent and identically distributed sample of $N_d$ events, $\mathcal{S} \equiv \{ d_i \}_{i=1}^{N_d}$. The unknown distribution from which the sample is drawn is denoted by $p(d)$ defined on the interval between the smallest and the largest measurable diameters, $[a,b]$. 
Our goal is to use GPR to infer $p(d)$ and its uncertainty from $\mathcal{S}$. 
The first step is to produce a frequency plot of the data with a suitable binning scheme, $\mathcal{B}$, defined by the bins' edges $b_i$ such that $\mathcal{B} = \{ b_i | b_1=a, b_{N_b+1}=b , b_{i+1}>b_i \}_{i=1}^{N_b+1}$. Let $N_b$ be the number of bins, $h_i$ the width of the $i^{\rm th}$ bin, and $n_i$ the number of drops within the $i^{\rm th}$ bin. The cumulative probability of a drop being in the first $i$ bins is approximated as 
\begin{equation}
    P_i  \equiv P(b_{i+1}) =  \int_{a}^{b_{i+1}} p(x) \textrm{d}x \approx  \frac{\sum_{j=1}^{i} n_j}{N_d (b_{i+1} - a)}   \ \ \ \textrm{for}\ \ \  i=1,\dots,N_b. \label{int_obs}
\end{equation} 

Hence, we compute the mean of the population, $\bar{d}$, which is an unbiased estimator of the first moment of the distribution $\mu_1$, yielding 
\begin{equation}
    \mu_1 \equiv \int_{a}^{b} p(x)x \textrm{d}x \approx \bar{d} = \frac{\sum_{i=1}^{N_d} d_i}{N_d}. \label{mean}
\end{equation}
The quantities $P_i$ and $\mu_1$ are linear functionals of $f(x)$, which we denote $L_i[p] \equiv P_i$, and $L_{N_f}[p] \equiv \mu_1$. 
The data set $\mathcal{D}$ is thus represented by (i) the definitions of these functionals, (ii) their observed values $\textbf{y}=[P_1,\dots,P_{N_b},\mu_1]$, and (iii) the variance of the noise corrupting these observations, $\mathbb{V}[\boldsymbol{\epsilon}]$. 

This data set is used to estimate the density $p$ at the test locations $\textbf{X}_*=[d_1,\dots,d_{N_*}]$ with the predictive equations \eqref{eq:pred1} to \eqref{eq:pred3}, where $\textbf{f}_* = p(\textbf{X}_*)$. To do so, the covariance between different measured features of $p$ is computed with equation \eqref{kL}. Specifically, the covariance between the observation of the $i^{\rm th}$ cumulative probability, $P_i$, and that of the first moment, $\mu_1$, is
\begin{equation}
    k_L(P_i, \mu_1) \equiv  L_i\left[ L_{N_f} \left[ k(\cdot ,\cdot) \right] \right] = L_{N_f}\left[ L_i \left[ k(\cdot ,\cdot) \right] \right] = \int_a^b \int_a^{b_{i+1}} k(z,z') z \textrm{d}z \textrm{d}z'. \label{cov1}
\end{equation}
Likewise, the covariance between the observations of $P_i$ and $P_j$ is
\begin{equation}
    k_L(P_i, P_j)  \equiv  L_i\left[ L_j \left[ k(\cdot ,\cdot) \right] \right] = L_j\left[ L_i \left[ k(\cdot ,\cdot) \right] \right] = \int_a^{b_{j+1}} \int_a^{b_{i+1}} k(z,z') \textrm{d}z \textrm{d}z'. \label{cov2}
\end{equation}
Finally, the covariance between the observations of $P_i$ and the function value at the $j^{\rm th}$ test point, $p(d_j)=L_{d_j}$, is
\begin{equation}
    k_L(P_i, L_{d_j}) \equiv L_i\left[ L_{d_j} \left[ k(\cdot ,\cdot) \right] \right] = L_{d_j}\left[ L_i \left[ k(\cdot ,\cdot) \right] \right] =  \int_a^{b_{i+1}} k(z,d_j) \textrm{d}z = \int_a^{b_{i+1}} k(d_j,z) \textrm{d}z. \label{cov3}
\end{equation}

\subsubsection{Non-stationary covariance function} \label{sec:nonstat_cov}
 We choose the specific form of the kernel $k=k(d,d')$ to compute the covariance in equations \eqref{cov1}, \eqref{cov2}, and \eqref{cov3} guided by the prior knowledge about the function we are learning. A number of studies have shown that the DSDs emerging from the breakup of ligaments and drop impacts can be approximated by Gamma \cite{marmottant2004a, marmottant2004b} as well as log-normal \cite{cohen1991} distributions. If the function $p(d)$ has similar shape to these distributions, we expect rapid variation of the density $p$ with the input $d$ around the peak, followed by a progressive transition to slower variation of the density towards the tail of the distribution (i.e., at large diameters). 
Consequently, the typical distance $\Delta d = |d-d'|$ over which the correlation between $p(d)$ and $p(d+\Delta d)$ decays, increases with the input, i.e., $\Delta d$ increases towards the tail of the distribution. 
A covariance function $k=k(d,d')$ with this property is said to be \emph{non-stationary}, as opposed to \emph{stationary} covariances, for which the correlation decays everywhere with the same law, $k(d,d') = k(\Delta d)$.
We allow the GPR model to capture this property of DSDs by employing the non-stationary version of the squared exponential covariance function proposed by \cite{gibbs1998}
\begin{equation}
    k(d,d') = \sigma_f^2 \left( \frac{2l_d(d) l_d(d')}{l_d^2(d) + l_d^2(d') } \right)^{1/2} {\rm exp}\left[ -\frac{ (d-d')^2 }{l_d^2(d) + l_d^2(d') } \right].  \label{kernel}
\end{equation}
The covariance in equation \eqref{kernel} is constructed so that the positive function $l_d(d)$ can be interpreted as the local correlation length, which is a function of the input $d$. Alternatively, for a given value of the parameter $\sigma_f$, the value of $l_d$ can be interpreted as the inverse of the characteristic gradient of $p$\footnote{As pointed out by Gibbs \cite{gibbs1998} (chapter $3.10.2$), who proposed the correlation function, this interpretation is not applicable for any arbitrary form of $l_d(d)$ due to the effect of the pre-factor to the exponential on the right-hand side of equation (\ref{kernel}). However, for the form of $l_c(d)$ used in this work, i.e., monotonic functions, see equation \eqref{ld}, this interpretation is accurate.}. Small values of $l_d$ imply that, locally, the modelled function varies significantly due to small variation of the input. This happens near the peak, where the model has high capacity, that is high ability to fit irregular data. Similarly, large values of $l_d$ imply that significant variations of the output require large variation of the input, i.e., in the tail the model has low capacity.  Finally, we note that the covariance function in equation \eqref{kernel} simplifies to the stationary squared exponential kernel if we let $l_d$ be constant, corresponding to a model that has uniform capacity with the input. Given the sharply peaked shape of DSD emerging from liquid atomization, a model with a stationary kernel (such as the squared exponential kernel) would either underfit the peak or overfit the tail of the distribution (results not shown). 

In the following, we consider two possible forms for $l_d$: A linear form, as it is the simpler form of non-uniformity, and the logarithmic form, to reproduce the nonlinear behaviour of log-normal and Gamma distributions, respectively
\begin{align}
    l_{\rm lin}(d) = a_0 + a_1 d , \hspace{4mm} {\rm and} \hspace{4mm} l_{\rm ln}(d) = b_0 + b_1 \ln (b_2d). \label{ld}
\end{align}
Consequently, the hyperparameters characterizing the kernel are either $\boldsymbol\theta \equiv \boldsymbol\theta_{\rm lin} = \{\sigma_f, a_0 , a_1\}$ or $\boldsymbol\theta \equiv \boldsymbol\theta_{\rm ln} = \{\sigma_f, b_0 , b_1 , b_2\}$, depending on whether we use $l_{\rm lin}(d)$ or $l_{\rm ln}(d)$, respectively. 
%

The noise that affects each measurement is assumed to be known and, in general,  $\mathbb{V}[\epsilon_i] \neq \mathbb{V}[\epsilon_j]$ for $i \neq j$. This scenario is relevant to real-world engineering and industrial applications, in which the accuracy of the experimental apparatus is a known function of the measured value, estimated during their calibration. In other words, we assume that the data points are obtained in a controlled environment with reliable estimates of their uncertainties. As a result, it is not necessary to either (i) treat the noise variance of each observation as an additional hyperparameter to be optimized during training (\emph{heteroscedasticity}), which would significantly increase the number of trained parameters; (ii) introduce the assumption that the variance is the same for all observations (\emph{homoscedasticity}). The latter assumption would be unrealistic in the present context due to the difficulties of measuring drops of very different sizes with the same accuracy, whether employing numerical or experimental techniques. 


The optimization problem in equation \eqref{training} for the training of the GPR is solved with a genetic algorithm \cite{conn1997globally}. Even though a gradient based approach would be computationally more efficient, our choice is justified by its simplicity of implementation and by the small computational load of the present problem, in which the number of data points used to perform the regression is $O(10)$ for each DSD. The integrals in equations \eqref{cov1}, \eqref{cov2}, and \eqref{cov3} are computed with a second-order finite-difference scheme by discretizing $k$ into $O(10^2)$ points. Each regression, such as those reported in the following, takes between $O(10^1)$ and $O(10^2)$ seconds to complete on 8 cores Intel i9 12900 processor. 

\subsection{Estimation of Drop Size Distribution} \label{sec:results} 
%

\begin{figure}[!htb] 
  \includegraphics[scale= 0.17]{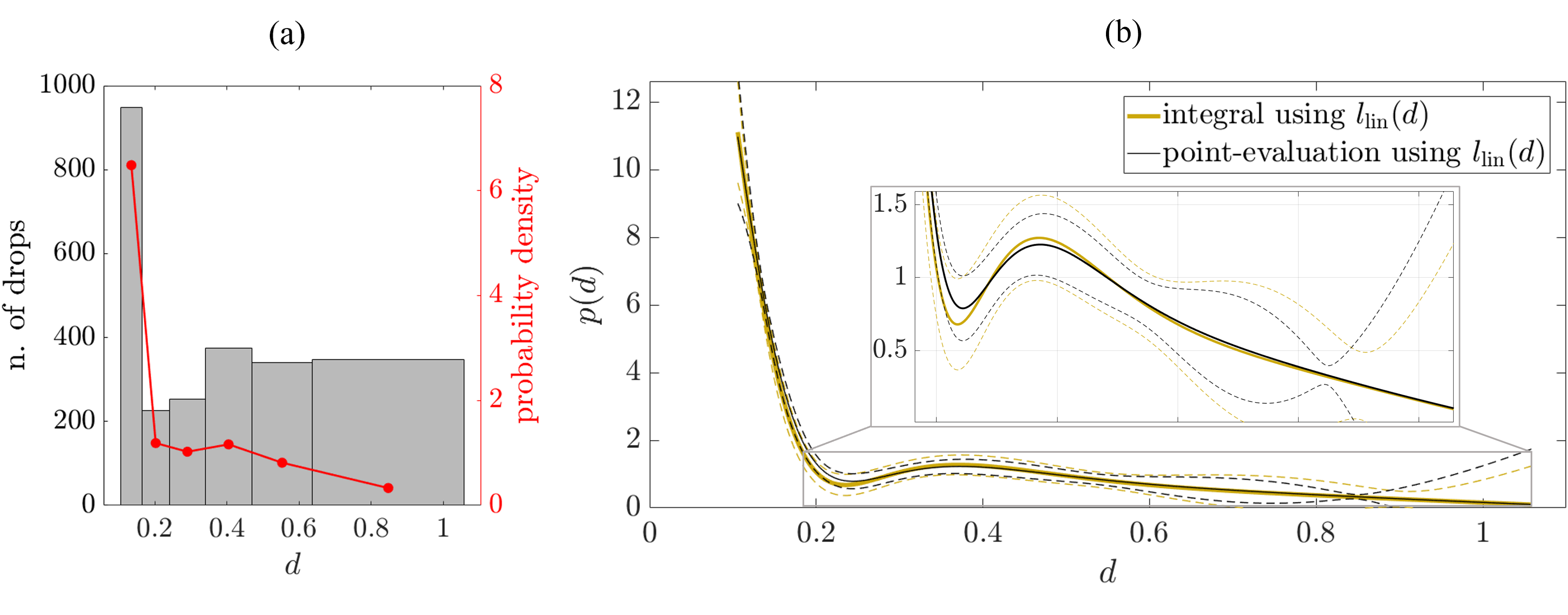}
  \caption{Data from the  numerical simulation with $\alpha=65$ deg, $Re = 20$, and $We=9$, see also figure \ref{fig:AMR} (d). (a) Histogram of the drops count in the simulated spray using 6 bins and associated probability (red circles). (b) DSD predicted by the GPR using observations of the mean and cumulative probabilities $p_{\rm int}$ (solid black lines), and point-like observations $p_{\rm pt}$ (dashed grey lines). The dashed lines represent the $95\%$ confidence intervals. The linear length scale $l_{\rm lin}(d)$ is employed to perform both regressions. }  \label{fig:DSD}
\end{figure}

The GPR framework discussed in section \ref{sec:features} is employed to estimate the DSD of a liquid spray. To evaluate the performance of the method, we refer to data relative to a spray characterized by $\alpha=65$ deg, $Re = 20$, and $We=9$ (see figure \ref{fig:AMR}). Upon filtering based on sphericity and size, the drops are sorted into bins, figure \ref{fig:DSD} (a). The corresponding probability density (red circles) is computed as $\pi_i \equiv n_i / (N_d h_i)$. The area under this line, up to each bin's edge, is equal to the cumulative probability $P_i$, defined in equation \eqref{int_obs}. The thick golden line in figure \ref{fig:DSD} (b) shows the posterior mean of $p_{\rm int}(d)$ inferred by the GPR by imposing the integrals $P_i$ and the value of the mean, $\mu_1$, computed according to equation \eqref{mean} (dashed lines are $95\%$ confidence intervals). Typically, the measurement accuracy is a function of the magnitude of the signal \cite{Geng2015}. To mimic this behaviour, in this example the variance of the additive Gaussian noise is set equal to $5\%$ of the measured value of each observation, that is $\mathbb{V}(\epsilon_i)=y_i \times 0.05$, except for the first two bins ($P_1$ and $P_2$), for which the error is assumed to be double. This reflects the fact that smaller drops are modelled less accurately, as their size approaches the spatial resolution of the solver.    


\subsubsection{Comparison between integral and point-like observations} \label{sec:comp_hist} 
An alternative method for inferring the DSD using GPR is to use traditional point-evaluation functionals as observations. These are the red circles corresponding to the probability density in figure \ref{fig:DSD} (a), which are treated as observations of the DSD at the center of each bin (i.e., the equivalent of interpolating the edges of a \emph{frequency polygon} \cite{Simonoff1996} with a GP model). By doing so, and by using the same covariance function and observations' noise variance, one obtains $p_{\rm pt}(d)$ (thin black curve in figure \ref{fig:DSD} (b)). 

The two methods give qualitatively similar results, however some important quantitative differences need be pointed out. First, we define the relative error between the population mean and the first moment of the reconstructed DSD, $e_{\mu} = 100|\bar{d} - \mu_1|/\bar{d}$. As reported in table \ref{Tab:err_int}, the expected value of $p_{\rm int}$ deviates by the mean diameter $\bar{d}$ by $0.27\%$, while that of $p_{\rm pt}$ by $3.0\%$, corresponding to a ten-fold reduction in the bias of the estimator $p_{\rm int}$. 
Second, the probability that a drop with diameter $b_i<d<b_{i+1}$ forms, i.e., the integrals
\begin{equation}
    p_i \equiv \int_{b_i}^{b_{i+1}} p(x) {\rm d}x, \label{int_bin}
\end{equation}
are compared with the associated quantity computed from the population of drops, that is $\pi_i$. The error in the prediction is quantified by $e_{i, {\rm int}}$ and $e_{i, {\rm pt}}$ 
computed as $e_i = 100| p_i - \pi_i |/\pi_i$, where $p=p_{\rm int}$ and $p=p_{\rm pt}$ in equation \eqref{int_bin} are used, respectively. The values of $e_{i, {\rm int}}$ and $e_{i, {\rm pt}}$ are listed in table \ref{Tab:err_int}. We observe that  the predictions provided by $p_{\rm int}$ are more accurate for all the bins  except 3 and 4, and that the error is reduced by up to two orders of magnitude in the tail of the distribution. The value of $p_i$ 
plays an important role as it is a direct indication of the volume (thus mass) of atomized fluid contained in drops with diameter $b_i<d<b_{i+1}$. As further discussed in section \ref{sec:concl}, this property of the DSD is key to practical applications.

\begin{table}[ht]
\caption{Data from the  numerical simulation with $\alpha = 65$ deg, $Re = 20$, and $We = 9$; drops sorted in $6$ bins, and $l_{\rm lin}(d)$ - see also figure \ref{fig:DSD}. Relative percentage error in the probability associated with each bin using integral observations, $e_{i, {\rm int}}$, and using point-evaluation observations, $e_{i, {\rm pt}} $. The last line reports the relative percentage error of the mean of the estimator.}
\centering
  \begin{tabular}{c c c}
    bin &  $e_{i, {\rm int}} [\%]$ & $e_{i, {\rm pt}} [\%]$ \\
    \hline 
    1       & 3.23 & 6.84 \\
    2       & 7.66 & 28.53\\
    3       & 2.63 & 2.15 \\
    4       & 5.39 & 2.41\\
    5       & 0.08 & 1.31\\
    6       & 0.63 & 3.25\\
    \hline
    $e_{\mu}$ & 0.27 & 3.0
  \end{tabular}
  \label{Tab:err_int}
\end{table}

\subsubsection{Effect of the kernel and binning scheme} \label{sec:kernel_effect} 
\begin{figure}[!htb] 
 \includegraphics[scale= 0.19]{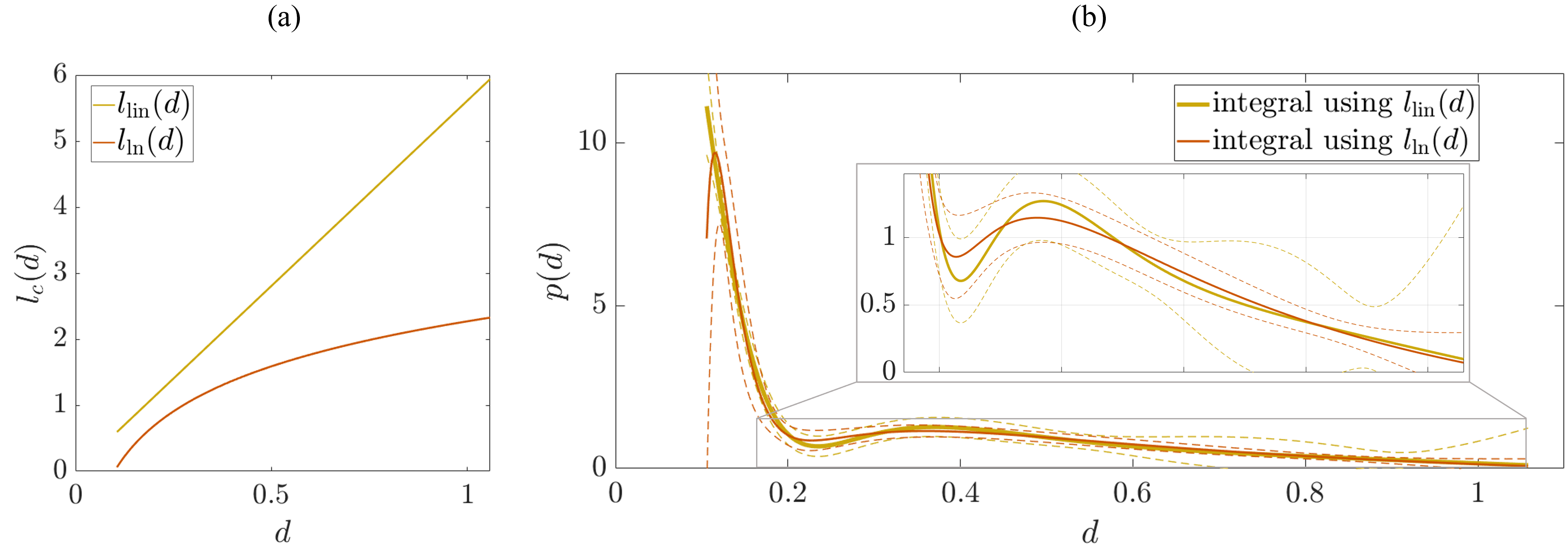}
  \caption{Data  from the  numerical simulation with $\alpha = 65$ deg, $Re = 20$, and $We = 9$; drops sorted in $6$ bins as shown in figure \ref{fig:DSD} (a). (a) Optimized characteristic length scales $l_{\rm lin}$ and $l_{\rm ln}$, corresponding to the GPR models in figure \ref{fig:DSD} (b) and \ref{fig:lc_DSD} (b), respectively. (b) DSD predicted by the GPR model using observations of the mean and cumulative probabilities $p_{\rm int}(d)$ employing the logarithmic length scale $l_{\rm ln}(d)$.}  \label{fig:lc_DSD}
\end{figure}

The choice of the kernel and the binning scheme are key steps in model selection as they affect the prediction. In this section, we discuss how the proposed approach allows us to mitigate the effect of these choices on the predicted DSD. 
The kernel defines the GPR model structure and encodes our assumptions about the functions that we are learning, which makes it a crucial ingredient in a GPR model \cite{Rasmussen2006}. Whereas it is clear from the context that our covariance function should be non-stationary, as discussed in section \ref{sec:nonstat_cov}, it is not obvious which specific form the kernel should have. We compare the predicted DSD obtained using the linear ($p_{\rm int, lin}$) and the logarithmic ($p_{\rm int, ln}$) length scale in figure \ref{fig:lc_DSD} (b). The two curves are obtained from the same observations as in figure \ref{fig:DSD}. The optimized lengths scales relative to $p_{\rm int}$ and $p_{\rm int, ln}$ are also plotted in figure \ref{fig:lc_DSD} (a). By comparing the two curves in figure \ref{fig:lc_DSD} (b), the choice of a linear form of the length scale removes the presence of the peak at small $d$, typical of lognormal and Gamma distributions, which are characterized by nonlinear decaying first derivatives. 
By design, the proposed method ensures that the bias of the estimator of the DSD (quantified by $e_{\mu}$) and the cumulative probabilities have similar accuracy in both cases, as shown by the comparison of the values of  $e_{i, {\rm int}}$ in table \ref{Tab:err_int} and table \ref{Tab:err_int_ln}. 
The fact that cumulative probabilities and bias of the DSD are robustly predicted against the choice of the kernel is a key feature of the proposed model.  

\begin{figure}[!htb] 
  \includegraphics[scale= 0.16]{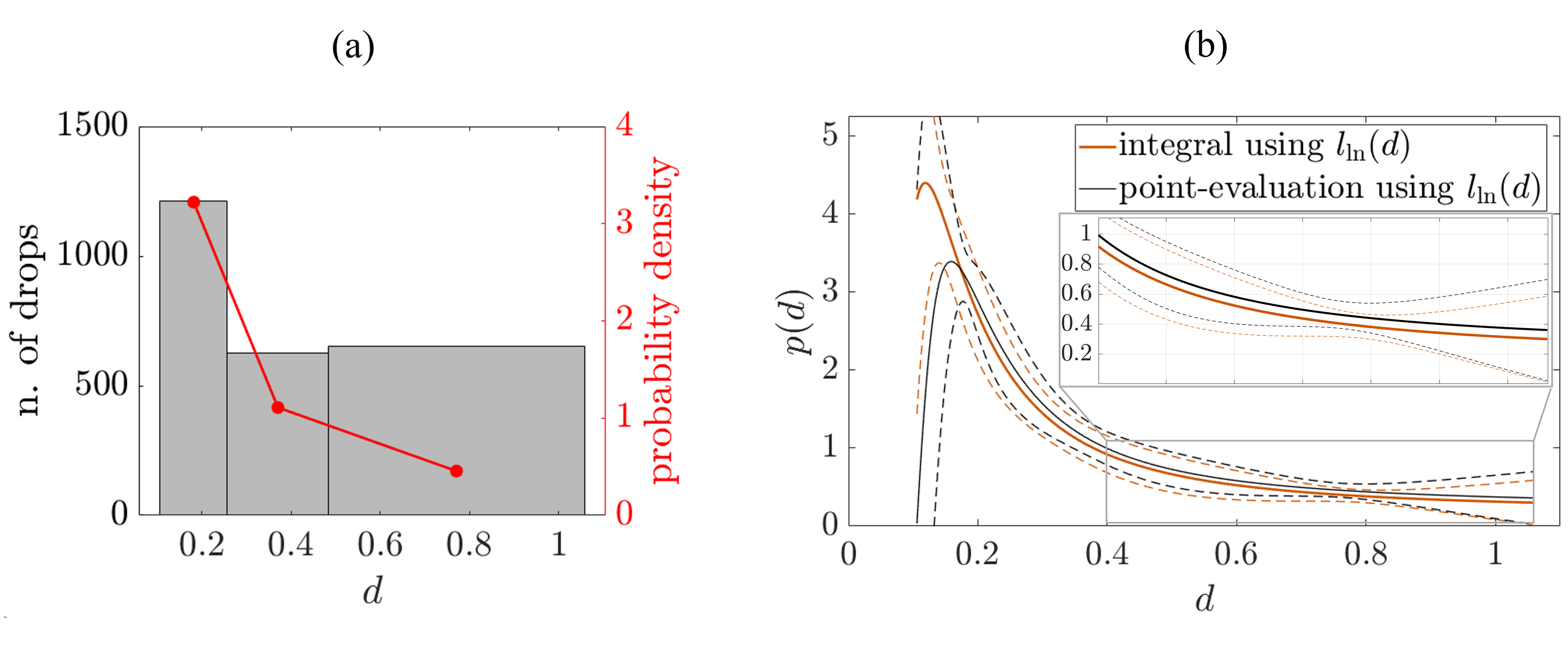}
  \caption{Data from the numerical simulation with $\alpha = 65$ deg, $Re = 20$, and $We = 9$. (a) Histogram of the drops count in the simulated spray using 3 bins and associated probability (red circles). (b) DSD predicted by the GPR using observations of the mean and cumulative probabilities $p_{\rm int}$ (solid green lines), and point-like observations $p_{\rm pt}$ (dashed grey lines). The dashed lines represent the $95\%$ confidence intervals. The logarithmic length scale $l_{\rm log}(d)$ is employed in both models. }  \label{fig:DSD_b3}
\end{figure}

When performing high-fidelity numerical simulations of atomization processes, e.g., \cite{ling2017spray}, small drops require high resolution to be resolved, and large drops are rare events, thus requiring the simulation to run for a long time to obtain a sufficient amount of realizations and for the statistics to converge. For this reason, obtaining large samples of drops is time consuming, thus the higher the fidelity of the simulation, the fewer the number of drops observed, and the coarser the binning scheme. 
We mimic this scenario in which fewer drops are observed by sorting the same population of drops in the examples above using a coarse binning scheme with only three bins. We train two models, one with integral and one with point-wise observations, using the resulting histogram reported in figure \ref{fig:DSD_b3} (a).
 The resulting DSDs predicted by the models are shown in figure \ref{fig:DSD_b3} (b). By comparing these DSDs with the red curve in figure \ref{fig:lc_DSD} (b) (obtained using the same kernel but more bins), we note that the details of the corresponding DSDs are different from those obtained using the more refined binning scheme. However, by inspection of table \ref{Tab:err_int_ln}, we observe that the magnitude of all the errors, $e_{i}$ and $e_{\mu}$, remains of the same order of magnitude 
 (see table \ref{Tab:err_int}) when integral observations are used. Finally, figure \ref{fig:DSD_b3} (b) shows that, when point-evaluation observations are used, the density estimator (i) is biased (last column of table \ref{Tab:err_int_ln}), (ii) fails to capture the magnitude of the peak, and (iii) overestimates the probability in the tail. 

\begin{table}[ht]
\caption{Relative percentage error in the probability associated with each bin, $e_{i, {\rm int}}$ and $e_{i, {\rm pt}}$, relative to the DSDs in figure \ref{fig:DSD_b3} (b) with $l_{\rm ln}(d)$. The last line reports the relative percentage error of the mean of the estimator.}
\centering  
  \begin{tabular}{c c c}
    bin &   $e_{i, {\rm int}} [\%]$ & $e_{i, {\rm pt}} [\%]$ \\
    \hline 
    1          & 0.67 & 13.20 \\
    2          & 2.68 & 9.31\\
    3          & 1.30 & 9.81 \\
    \hline
    $e_{\mu}$  & 0.57 & 6.27  
  \end{tabular} 
  \label{Tab:err_int_ln}
\end{table}

\section{GPR-based surrogate model of a liquid spray}   \label{sec4}

In this section, we use the method for the reconstruction of the DSD introduced in section \ref{sec2} and tested in section \ref{sec:GPR_pdf} to develop a data-driven surrogate model (SM) of a flat fan spray. The structure of the SM is introduced in section \ref{sec:surr_mod_str}, and the predictions are discussed in \ref{sec:surr_mod_res}.

\subsection{Structure and training of the surrogate model} \label{sec:surr_mod_str} 
\begin{figure}[!htb] 
  \includegraphics[scale= 0.43]{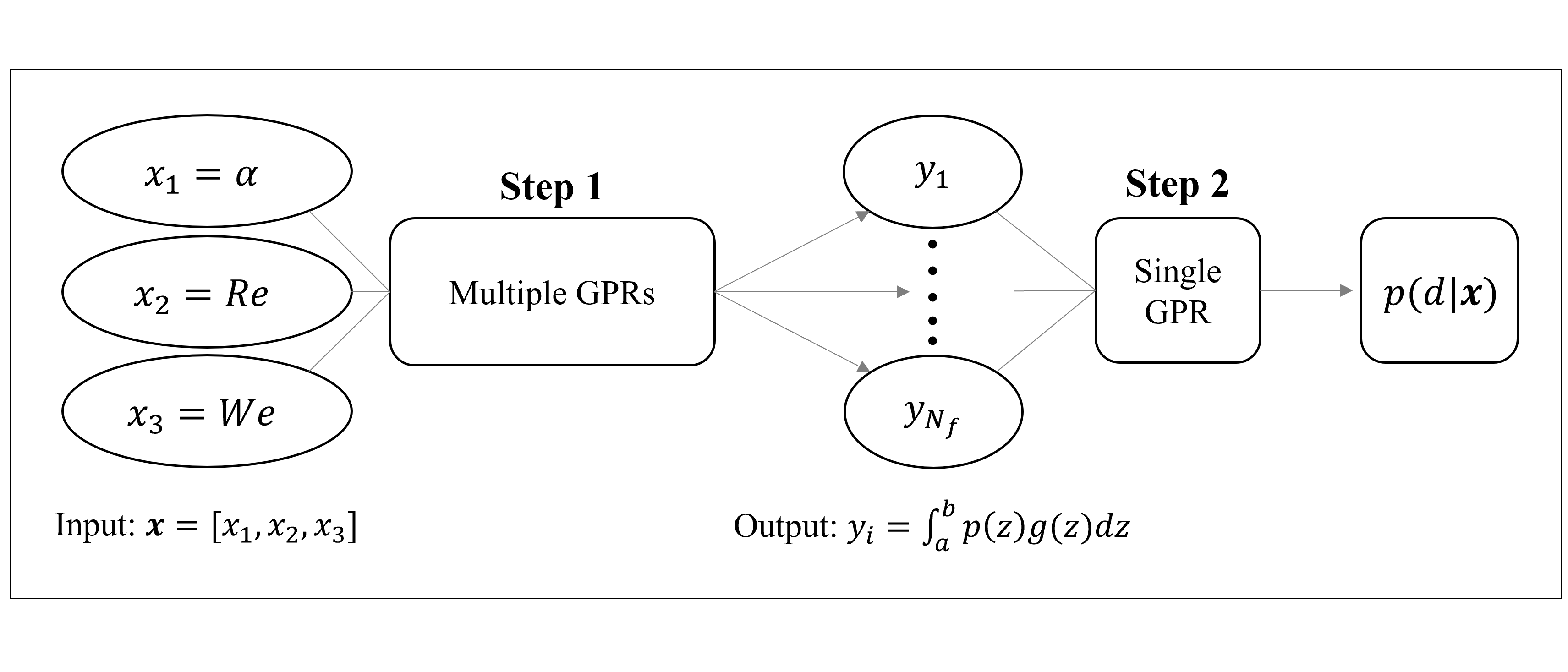}
  \caption{Schematic of the structure of the surrogate model that maps each parameter, $\textbf{x}=[x_1,x_2,x_3]$, to the predicted DSD $p(d|\textbf{x})$. The first step predicts how each feature $y_i$ of the DSD changes with the input by training multiple GPR models, one for each feature. In the second step the predicted features are used as data to reconstruct the DSD with one final GPR model, which uses integral observations.}  \label{fig:2stepalg}
\end{figure}

\begin{figure}[!htb] 
  \includegraphics[scale= 0.18]{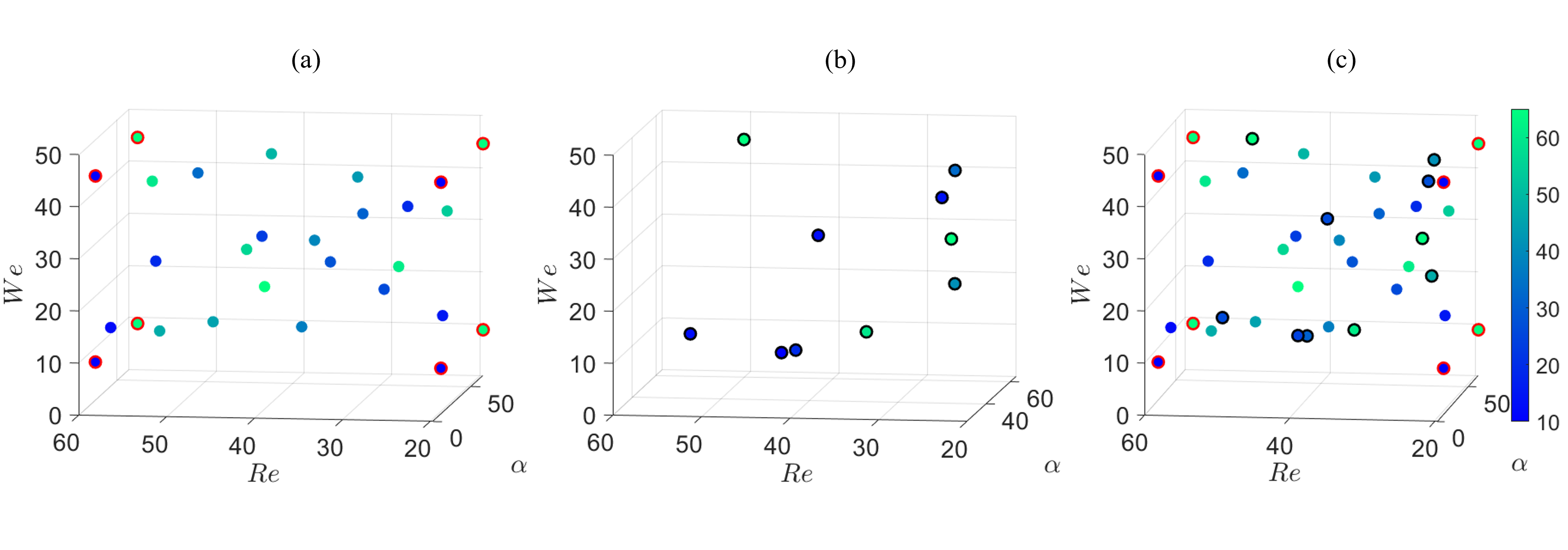}
  \caption{Design of experiments for the training data. (a) First batch of numerical simulations as points in the input space obtained from a Latin hypercube design and the corners of the cube (red circles). (b) Second batch of $10$ simulations identified by minimizing the posterior signal-to-noise ratio. (c) Entire set of simulations performed. The colorbar refers to the value of the spray angle $\alpha$. }  \label{fig:DOE}
\end{figure}

The SM is designed to predict the DSD of the spray, $p(d|\textbf{x})$, given the control/design parameters $\textbf{x} = [\alpha,\ Re,\ We]$. This comprises two main steps, as depicted in figure \ref{fig:2stepalg}. 
First, in Step 1, the SM learns how the features of the DSD, $\textbf{y} = [y_1, ..., y_{N_f}]$, change with $\textbf{x}$. Then, their predicted value at a test location, $\textbf{y}(\textbf{x}_*)$, and the associated uncertainty, $\mathbb{V}(\boldsymbol\epsilon)$, are used to reconstruct the DSD using the method described in section \ref{sec:GPR_pdf} (i.e., Step 2).
Each of the $N_f$ GPR models involved in Step 1 is trained using data from $38$ numerical simulations spanning the input space $\mathcal{X} = [10, 65] \times [20, 59] \times [9, 45]$, figure \ref{fig:DOE} (c). Specifically, we first run a batch of $28$ numerical experiments, of which $20$ according to a Latin hypercube sampling and $8$ simulations at the edges of the three-dimensional input space, see figure \ref{fig:DOE} (a). At this point, $N_f$ GPR models are trained to learn each feature $y_i$ based on available data. These models are used to design a second batch of $10$ additional numerical experiments, figure \ref{fig:DOE} (b). The criterion used for the DOE is the minimization of the signal-to-variance ratio of the predicted $y_i(\textbf{x})$ within the interrogation region $\mathcal{X}$. More details on this iterative GPR-based method for the optimal design of experiments can be found in the work by Geng et al. \citep{Geng2015}.  

Within Step 1, the $i^{\rm th}$ GPR model is trained using the squared exponential kernel $\tilde{k}_i$ for $i=1,\dots, N_f$ with automatic relevance determination (ARD) \citep{Neal_ARD}. The kernel is defined as
\begin{equation}
    \tilde{k}_i(\textbf{x}_m, \textbf{x}_n) = \tilde{\sigma}_{f,i}^2 \exp\left(-\frac{1}{2} (\textbf{x}_m-\textbf{x}_n)^{\textrm{T}} M_i (\textbf{x}_m-\textbf{x}_n) \right) \label{ARD_SE}
\end{equation}
with hyperparameters $\boldsymbol\theta_i = \{M_i, \tilde{\sigma}_{f,i}^2\}$, where $M_i=\textrm{diag}(\textbf{l}_i)^{-2}$, and $ \textbf{l}_i = [l_{i}^{(\alpha)}, l_{i}^{(\rm Re)}, l_{i}^{(\rm We)}]$. Physically, each element of $\textbf{l}_i$ plays the role of a characteristic length scale indicating the degree to which a change in the associated design parameter (i.e. $\alpha$, $Re$, $We$) is necessary to induce a significant change in the $i^{\rm th}$ feature. To ease the comparison between length scales, the input space is scaled before training so that all data points are contained in $\mathcal{X}' = [0, 1] \times [0, 1] \times [0, 1]$ -- also referred to as \emph{min-max} normalization \citep{LI2011256}.

\subsection{Results} \label{sec:surr_mod_res} 

\subsubsection{Step 1: Performance and sensitivity to design parameters} \label{sec:SM_general} 

Table \ref{Tab:lc} reports the values of the characteristic length scales associated with the $i^{\rm th}$ feature, $\textbf{l}_i$. 
We observe that features are more sensitive to changes in $We$, which, on average, is associated with smaller values of $l_i^{(\rm We)}$, followed by the spray angle $\alpha$ and then $Re$. 
In particular, the average diameter decreases as $We$ increases, as visible by comparing figures \ref{fig:AMR} (c-d) with (e-f), but it is virtually unaffected by $Re$, as suggested by the large value of $l_9^{(\rm Re)}$. 
The performance of each of the $N_f$ GPR models involved in Step 1 is evaluated by performing leave-one-out and 2-fold cross-validation, and by computing the associated normalized mean squared error, ${\rm MSE}_{\rm LOO}$ and ${\rm MSE}_{\rm 2-fold}$, respectively. Their values, expressed in percentages of the mean predicted value, are reported in the last two columns of table \ref{Tab:lc}. The average of ${\rm MSE}_{\rm LOO}$ amongst all the $N_f$ features is $1.25 \%$, showing that the models fit the data well. Notably, the models continue to perform well when tested by 2-fold cross validation, i.e., when half of the data points are used for training and half for testing, which result in an average value of ${\rm MSE}_{\rm 2-fold}$ approximately $1.89 \%$. These observations not only demonstrate that the $N_f$ GPR models provide accurate predictions using only $19$ randomly sampled data points, but also that they do not suffer over-fitting when all the $38$ simulations are used for training.


%
\begin{table}[ht]
\caption{Characteristic length scales, $l_i^{(\alpha)}$, $l_i^{(\rm Re)}$, and $l_i^{(\rm We)}$, associated with each feature, $p_i$ and $\mu_1$, along the dimension of the normalized design space $\mathcal{X}'$. Extremes of the bin associated with the probabilities $p_i$, $b_i$ and $b_{i+1}$. Mean squared error expressed as percentage of the mean predicted value based on leave-one-out and 2-fold cross-validations, ${\rm MSE}_{\rm LOO}$ and ${\rm MSE}_{\rm 2-fold}$, respectively. Comparing $l_i^{(\alpha)}$, $l_i^{(\rm Re)}$, and $l_i^{(\rm We)}$ we note that $We$ is the most influential parameter and that $Re$ has a negligible effect on the mean drop size. The similar values of ${\rm MSE}_{\rm 2-fold}$ to ${\rm MSE}_{\rm LOO}$ shows that performances do not deteriorate substantially when the training data set is halved.}
\centering  
  \begin{tabular}{c c c c c c c c c c c}
    Feature & \hspace{.4cm} &  $l_i^{(\alpha)}$ & $l_i^{(\rm Re)}$ & $l_i^{(\rm We)}$ & \hspace{.4cm} & $b_i$ & $b_{i+1}$ & \hspace{.4cm} & ${\rm MSE}_{\rm LOO}$ $[\%]$ & ${\rm MSE}_{\rm 2-fold}$ $[\%]$ \\
    \hline 
    $L_1=p_1$  &   & 2.8 & 29.7 & 9.0 & & 0.085   & 0.146 & &  0.06 & 0.05\\
    $L_2=p_2$  &   & 5.1 & 14.2 & 2.0 & & 0.146   & 0.213 & &  0.43 & 0.58\\
    $L_3=p_3$  &   & 9.1 & 5.7 & 1.6  & & 0.213   & 0.287 & &  0.35 & 0.41\\
    $L_4=p_4$  &   & 4.0 & 6.1 & 1.9  & & 0.287   & 0.368 & &  0.73 & 0.93 \\
    $L_5=p_5$  &   & 6.8 & 23.0 & 3.2 & & 0.368  & 0.457  & &  0.91 & 0.96\\
    $L_6=p_6$  &   & 2.1 & 18.0 & 2.3 & & 0.457  & 0.555  & &  1.66 & 7.18\\
    $L_7=p_7$  &   & 1.6 & 7.9 & 1.3  & & 0.555   & 0.663 & &  1.85 & 2.23\\
    $L_8=p_8$  &   & 2.3 & 14.1 & 0.9 & & 0.663  & 0.794  & &  5.20 & 4.46\\
    $L_9=\mu_1$&   & 3.3 & 71.6 & 2.6 & &         &       & &  0.05 & 0.15\\
  \end{tabular} 
  \label{Tab:lc}
\end{table}

\subsubsection{Step 2: Exploring the design space} \label{sec:SM_Des_par}  
\begin{figure}[!htb] 
  \includegraphics[scale= 0.18]{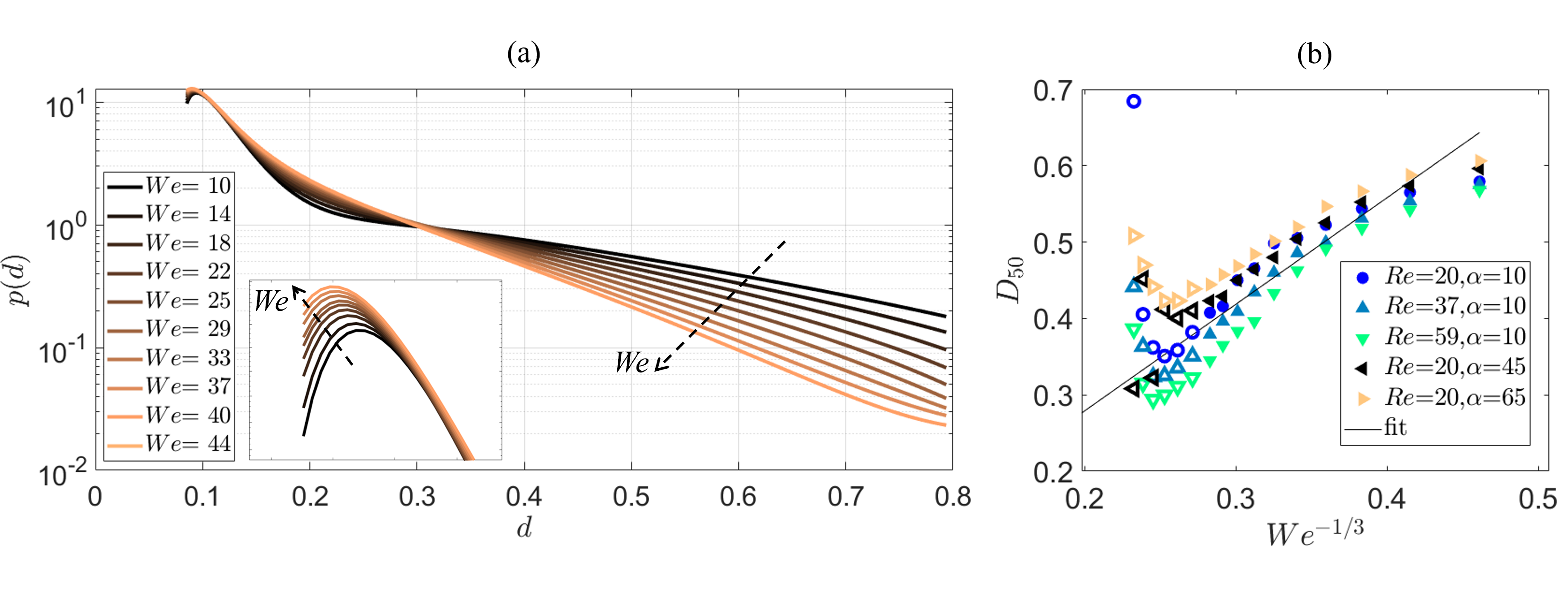}
  \caption{ (a) Drop size distributions predicted by the SM at different values of the Weber number (inset: close up of the peaks). (b) Volume mean diameter $D_{50}$ as a function of $We^{-1/3}$ for different spray angles and Reynolds numbers. Open symbols are in extrapolation region. }  \label{fig:We_D50}
\end{figure}

\begin{figure}[!htb] 
  \includegraphics[scale= 0.3]{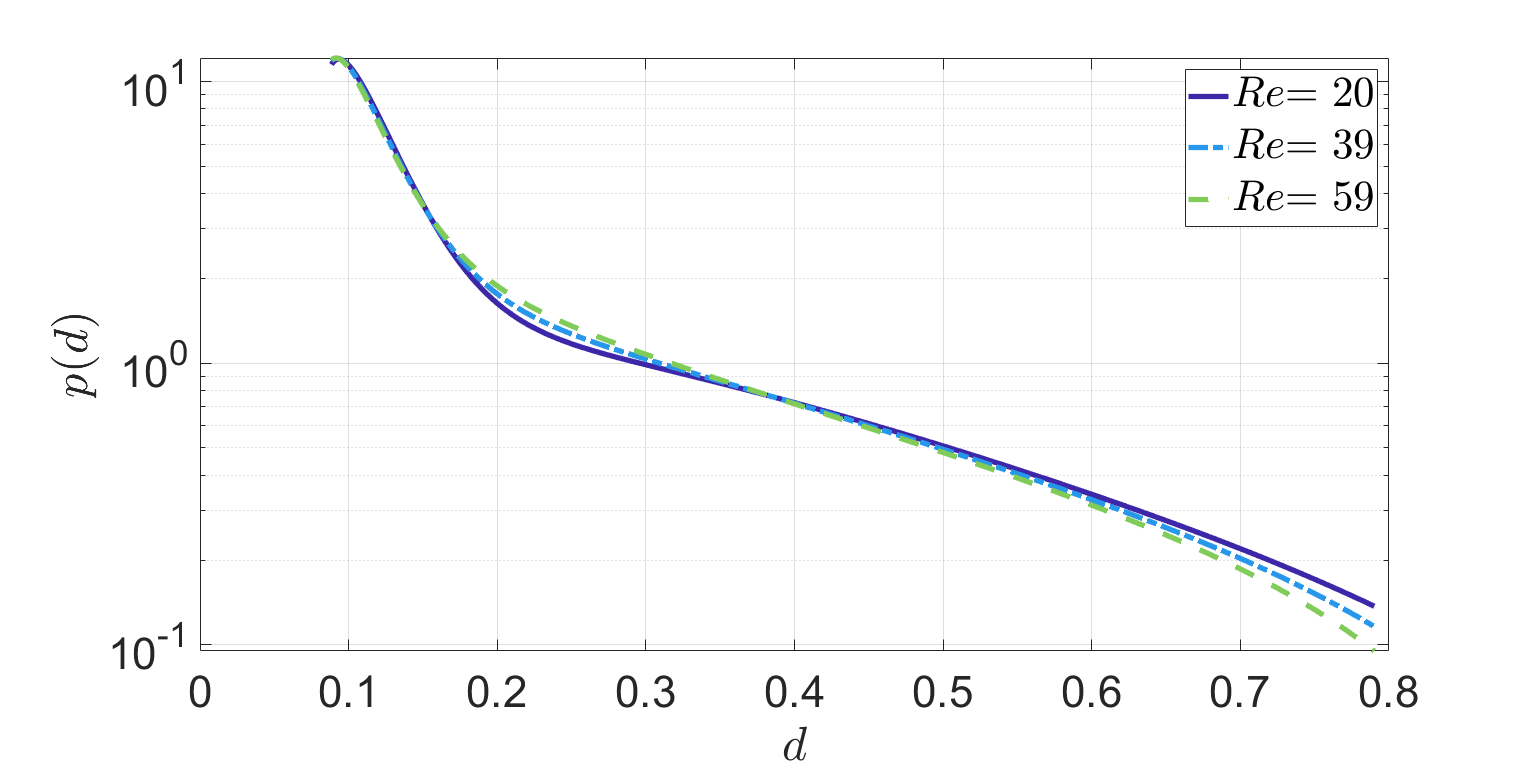}
  \caption{ Drop size distributions predicted by the SM at different values of the Reynolds number. }  \label{fig:Re3}
\end{figure}

\begin{figure}[!htb] 
  \includegraphics[scale= 0.18]{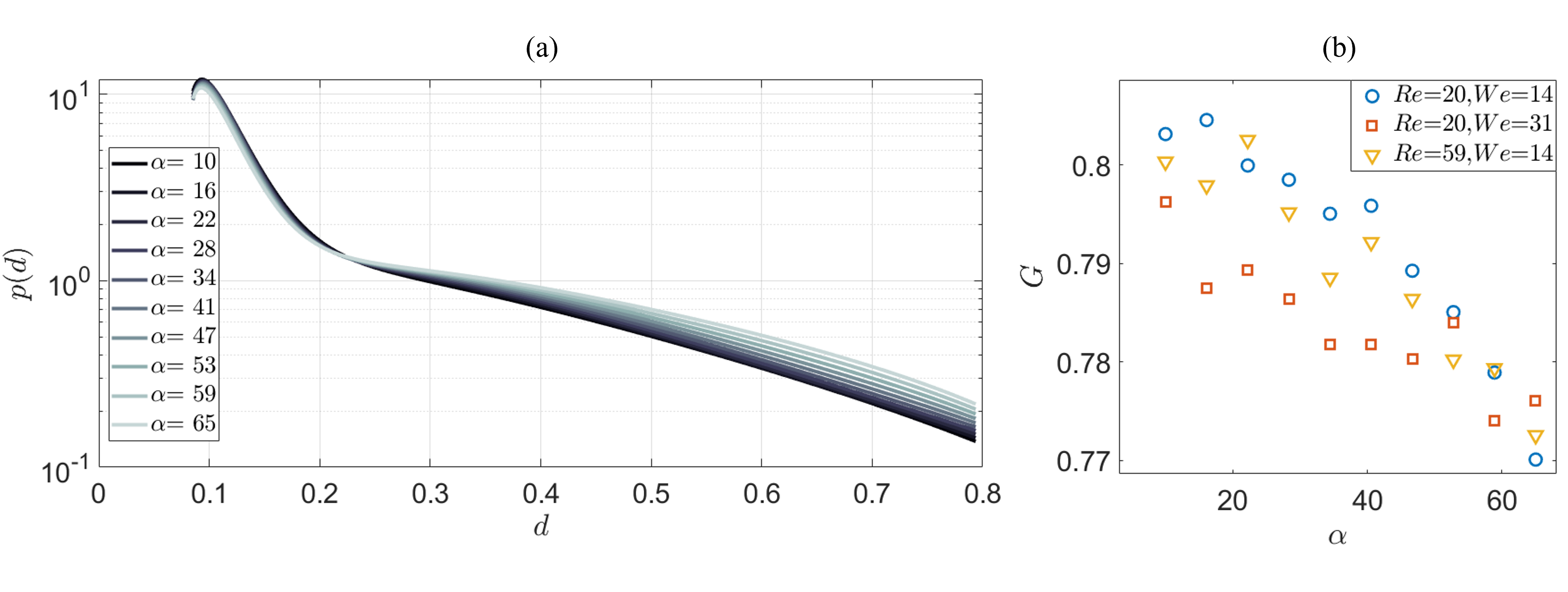}
  \caption{  (a) Drop size distributions predicted by the SM at different values of the spray angle $\alpha$. (b) Gini index of the volume distribution as a function of $\alpha$. }  \label{fig:alp_Gini}
\end{figure}

By combining the two steps in figure \ref{fig:2stepalg}, we explore the design space and produce DSDs at different values of $We$, $Re$, and  $\alpha$, as reported in panel (a) of figures \ref{fig:We_D50},  \ref{fig:Re3}, and \ref{fig:alp_Gini}, respectively. 
Each DSD has its maximum near the minimum diameter resolved in the numerical simulations, followed by a decrease of the probability with increasing drop diameter. This leads to a plateau-like region of intermediate-sized drops where the probability tends to decrease less steeply, followed by the tail. As expected from considerations of the instability of the liquid sheet leading to ligament formation \cite{Villermaux2007,villermaux2002life}, increasing $We$ leads to a faster decay of the DSD tail and a corresponding increase in the probability at small and intermediate diameters. The reduction in $We$ also shifts the peak of the DSD towards larger diameters, as shown in the inset of figure \ref{fig:We_D50} (a), in agreement with experimental evidence \cite{Kooij2018}. We remark that the SM can qualitatively capture such a shift in the peak despite the fact that it is located in a low-confidence region (i.e., small drops) and that the same binning scheme, reported in table \ref{Tab:lc}, has been used for all the simulations. 

The scaling $D_{50} \sim We^{-1/3}$ of the volume mean diameter $D_{50}$ predicted by Kooij et al. \cite{Kooij2018} is also tested for different values of the Reynolds number and the spray angle, see figure \ref{fig:We_D50} (b). The values of $D_{50}$ are computed from synthetic populations of drops generated via a pseudo-random number generator from the DSDs predicted by the SM. Towards the right side of the plot the trend starts to be nonlinear which could be signalling a change on the relative importance of different breakup mechanisms at due to the small values of the Weber numbers, which are several order of magnitude smaller than those for which the scaling was observed. On the left side of the figure are reported the predicted $D_{50}$ in the extrapolation region, i.e., for $We>45$ (empty symbols). Our results show that the SM manages to predict the correct scaling up to $We^{-1/3} \approx 0.25$, thus corresponding to a $40 \%$ increase with respect to the highest $We$ in the training data. 
Qualitatively, both $D_{50}$ and the characteristics of the DSD appear to be affected in a similar way by the Reynolds number, even if it has a quantitative smaller effect (figure \ref{fig:Re3}). 

Finally, we observe that the main effect of increasing the spray angle is to increase the probability of drops of intermediate size up until the tail of the distribution, as shown in figure \ref{fig:alp_Gini} (a). To better interpret this result, we use the DSDs to produce equally-sized synthetic populations of $N_d = O(10^4)$ drops, $\mathcal{S} = \{ d_i \}_{i=1}^{N_d}$, for different values of $\alpha$, $Re$ and $We$. Then each diameter $d_i$ is converted into the equivalent volumes $v_i = \pi d_i^3 / 6$, and the populations of volumes are finally used to compute the Gini index
\begin{equation}
    G = \frac{\sum_{i=1}^{N_d} \sum_{j=1}^{N_d} |v_i - v_j| }{ 2 N_d^2 \bar{v}} \label{gini_ind},
\end{equation}
where $\bar{v}$ is the arithmetic mean of the population of volumes.  
The Gini index, originally developed to measure the income inequality within a given social group \cite{weymark1981Gini}, is used here to measure, within a given spray, the inequality in volume (i.e., mass) distribution of drops. A Gini index of one corresponds to a spray in which all the volume is in one single drop, while a Gini index of zero corresponds to the case where the volume of the spray is distributed in equally sized drops. Our results, reported in figure \ref{fig:alp_Gini} (b), show that a larger spray angles correspond to smaller $G$, thus to more equally distributed populations in terms of the volume in each drop\footnote{The same analysis could be performed in term of linear size, i.e., diameters $d_i$, in which case we could compute $G$ directly from the continuous DSD. We choose to use the volume as directly related to the mass distribution, thus more practically relevant.}. This result can be explained by noting that the drops formed at the rims of the spray are larger than those generated by the flapping instability of the liquid sheet \cite{villermaux2002life,Kooij2018}, see figure \ref{fig:AMR} (c) and (d). The relative importance of these larger drops increases in narrow sprays, leading to a more unequal distribution of the volume and, thus, larger $G$.

%


\section{Conclusions} \label{sec:concl}

We propose a quantitatively accurate data-driven method to predict the drop size distribution (DSD) of a flat fan spray as a function of the spray angle and  working conditions, which are parameterized by the Reynolds and Weber numbers of the jet. The method is based on Gaussian Process Regression (GPR) models,  therefore, the prediction has confidence intervals, and takes into account the uncertainty of the data. Key to our method is the observation of integrals (i.e., cumulative probabilities) and the first moment (i.e., expected value) to reconstruct the unknown probability  density function. These features of the DSD determine the percentage of atomized fluid contained in droplets of a certain size and the bias of the estimator. 
First, we show that integral observations for the inference of the continuous DSD provides physically consistent and accurate predictions, which overcome the limitations of the commonly used point estimates of the DSD.
Second, we show that the training and predictions of the proposed model are robust with respect to the selection of the numerical simulations used for training and test, and the model's hyperparameters.   
Third, we propose kernels, which are physically motivated. 
%
%
The DSD is estimated without assuming its shape a priori, such as that of Gamma or log-normal distributions, as typically performed in the literature. We embed the physical fact that liquid atomization gives rise to  sharply peaked, heavy-tailed distribution into the choice of the covariance function. 
Our analysis demonstrates that the SM is versatile for the prediction of more intricate (e.g., bimodal) distributions, whilst retaining the sharply peaked, heavy-tailed characteristics.

 The proposed SM can be adapted to characterize other complex processes, the outcome of which is a probability density function  that changes with the control parameters. For example, these include other multiphase problems, such as the modelling of the bubble size distribution in chemical reactors. 
The application of this approach to experimental data and larger parameter spaces is the subject of current and future work.

%
%
\section*{Acknowledgments}
We acknowledge funding from the Engineering and Physical Sciences Research Council, UK, through the Programme Grant PREMIERE (EP/T000414/1). L.M. also acknowledges financial support from the ERC Starting Grant PhyCo 949388 and the UKRI AI for Net Zero grant  EP/Y005619/1.

%


\bibliography{My_Collection_1}

\end{document}